\documentclass[twocolumn,showpacs,aps,floatfix,prd,nofootinbib]{revtex4}
\usepackage{graphicx}
\usepackage{dcolumn}
\usepackage{amssymb}
\usepackage{amsmath}
\usepackage{colordvi}
\usepackage{color}
\usepackage{hhline}

\RequirePackage{xspace}
\newcommand{\BABARPubYear}    {09}
\newcommand{\BABARPubNumber}  {034}
\newcommand{\SLACPubNumber} {13953}

\usepackage{relsize}
\def\babar{\mbox{\slshape B\kern-0.1em{\smaller A}\kern-0.1em
    B\kern-0.1em{\smaller A\kern-0.2em R}}}

\def\Y#1S{\ensuremath{\Upsilon{(#1S)}}\xspace}

\def\pep2{PEP-II}

\long\def\inst#1{\par\nobreak\kern 4pt\nobreak
  {\it #1}\par\vskip 10pt plus 3pt minus 3pt}
  
\begin{document}

\begin{flushleft}
SLAC-PUB-\SLACPubNumber \\
\babar-PUB-\BABARPubYear/\BABARPubNumber \\
\end{flushleft}

\title{
Measurement of the $\gamma\gamma^\ast \to \eta_c$ 
transition form factor}

\author{J.~P.~Lees}
\author{V.~Poireau}
\author{E.~Prencipe}
\author{V.~Tisserand}
\affiliation{Laboratoire d'Annecy-le-Vieux de Physique des Particules (LAPP), Universit\'e de Savoie, CNRS/IN2P3,  F-74941 Annecy-Le-Vieux, France}
\author{J.~Garra~Tico}
\author{E.~Grauges}
\affiliation{Universitat de Barcelona, Facultat de Fisica, Departament ECM, E-08028 Barcelona, Spain }
\author{M.~Martinelli$^{ab}$}
\author{A.~Palano$^{ab}$ }
\author{M.~Pappagallo$^{ab}$ }
\affiliation{INFN Sezione di Bari$^{a}$; Dipartimento di Fisica, Universit\`a di Bari$^{b}$, I-70126 Bari, Italy }
\author{G.~Eigen}
\author{B.~Stugu}
\author{L.~Sun}
\affiliation{University of Bergen, Institute of Physics, N-5007 Bergen, Norway }
% ...
\collaboration{The \babar\ Collaboration}
\noaffiliation

\begin{abstract}
We study the reaction $e^+e^-\to e^+e^-\eta_c,\; \eta_c\to K_S K^\pm\pi^\mp$
and obtain $\eta_c$ mass and width values $2982.2\pm0.4\pm1.6$ MeV/$c^2$
and $31.7\pm1.2\pm 0.8$ MeV, respectively. We find
$\Gamma(\eta_c\to\gamma\gamma){\cal B}(\eta_c\to
K\bar{K}\pi)=0.374\pm0.009\pm0.031$
keV, and measure
the $\gamma\gamma^\ast \to \eta_c$ transition form factor
in the momentum transfer range from 2 to 50 GeV$^2$. The analysis is
based on 469 fb$^{-1}$ of integrated luminosity collected at \pep2\ with
the \babar\ detector at $e^+e^-$ center-of-mass energies near 10.6 GeV.
\end{abstract}

\pacs{14.40.Lb, 13.40.Gp, 12.38.Qk}

\maketitle
\setcounter{footnote}{0}

\section{Introduction}\label{intro}
\begin{figure}
\includegraphics[width=.33\textwidth]{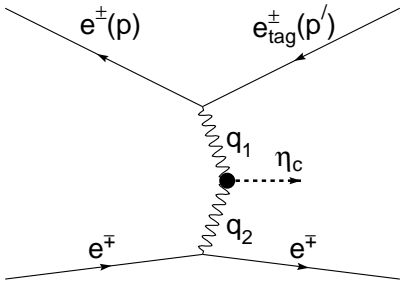}
\caption{The Feynman diagram for the $e^+e^-\to e^+e^-\eta_c$ two-photon
production process.
\label{fig1}}
\end{figure}

In this paper we study the process  
\begin{equation}
e^+e^-\to e^+e^-\eta_c,
\end{equation}
where the $\eta_c$ meson ($J^{PC}=0^{-+}$),
the lowest lying charmonium state,
is produced via the two-photon production
mechanism illustrated in Fig.~\ref{fig1}.
We measure the differential cross section for this process in the single-tag
mode where one of the outgoing electrons\footnote{Unless otherwise
specified, we use the term ``electron'' for either an electron or a positron.}
is detected (tagged), while the other electron  is scattered at a small angle
and hence is undetected (untagged). The tagged electron emits
a highly off-shell photon with squared momentum transfer 
$q^2_1 \equiv -Q^2 = (p^\prime-p)^2$, where $p$ and $p^\prime$ are the 
four-momenta of the initial- and final-state electrons. 
The momentum transfer squared to the untagged electron ($q^2_2$) is near zero. 
The differential cross section $d\sigma(e^+e^-\to e^+e^-P)/dQ^2$ for 
pseudoscalar meson ($P$) production depends
on only one form factor $F(Q^2)$, which describes
the $\gamma\gamma^\ast \to P$ transition. To relate the differential
cross section to the transition form factor we use the formulae
for the $e^+e^-\to e^+e^-\pi^0$ cross section in Eqs.~(2.1) and (4.5) of
Ref.~\cite{BKT}.

According to perturbative QCD (pQCD),
the transition form factor can be presented as a convolution of a
calculable hard scattering amplitude for $\gamma\gamma^*\to c\bar{c}$
with a nonperturbative light-cone wave function $\phi_{\eta_c}$~\cite{LB}.
The measurement of the form factor allows us to test the pQCD prediction
and to obtain information on the shape of the $\eta_c$ wave function.
The $Q^2$ dependence of the form factor is
studied theoretically  in Refs.~\cite{th1,th2} using pQCD, and
in Ref.~\cite{lqcd} using the lattice QCD approach.

The $\eta_c$ transition form factor was measured by the L3
Collaboration~\cite{L3} with very small data sample.
In this paper we study the $e^+e^-\to e^+e^-\eta_c$ reaction
in the $Q^2$ range from 2 to 50 GeV$^2$.
The $\eta_c$ is observed via the $\eta_c \to K_S K^\pm\pi^\mp$ 
decay,\footnote{The use of charge
conjugate reactions is implied throughout unless explicitly stated otherwise.}
which 
allows the $\eta_c$ to be selected with relatively low background.
The $\eta_c$ two-photon width and branching fractions are not well 
measured~\cite{pdg}. 
Therefore, we also study no-tag data ($Q^2\sim 0$),
measure the product $\Gamma(\eta_c\to\gamma\gamma){\cal B}(\eta_c\to K\bar{K}\pi)$, 
and normalize the transition form factor $F(Q^2)$ to $F(0)$.
The measured values of the $\eta_c$ mass and width obtained in different 
experiments have
a large spread~\cite{pdg}. The high statistics no-tag data sample allows us to
extract precise values for these parameters.

\section{The \babar\ detector and data samples}
\label{detector}
We analyze a data sample corresponding to an integrated
luminosity of 469~fb$^{-1}$ recorded with
the \babar\ detector~\cite{babar-nim} at the \pep2\ 
asymmetric-energy storage rings at SLAC. At \pep2, 9 GeV electrons collide with
3.1 GeV positrons to yield an $e^+e^-$ center-of-mass (c.m.) energy of 10.58~GeV
(the $\Upsilon$(4S) resonance). 
About 10\% of the data used in the present analysis were 
recorded about 40 MeV below the resonance.
\begin{figure*}
\includegraphics[width=.4\textwidth]{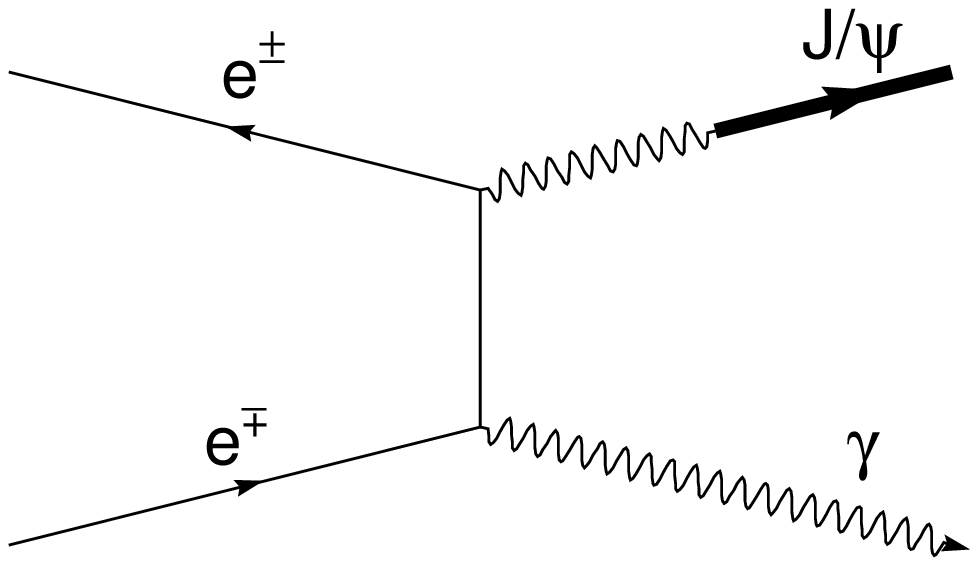}
\hfill
\includegraphics[width=.4\textwidth]{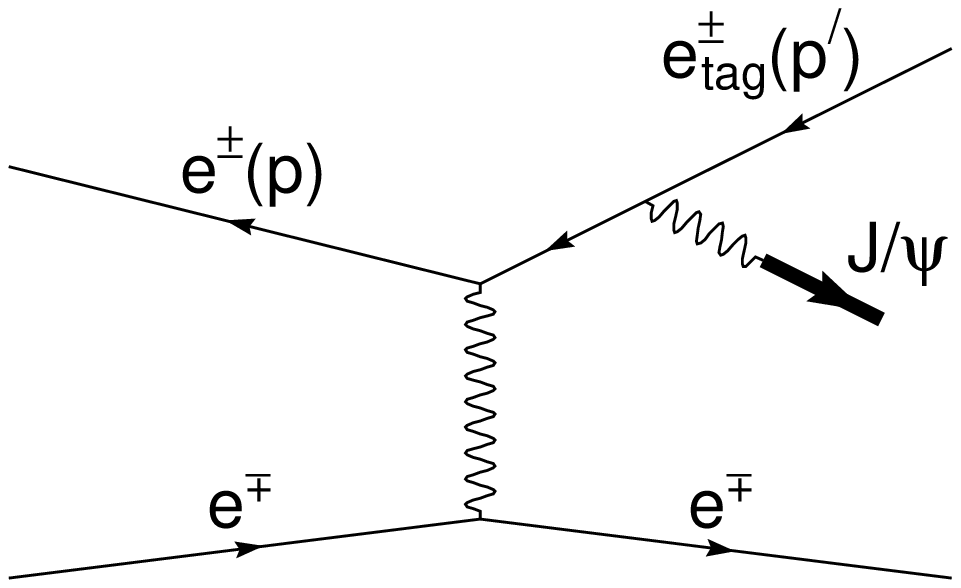}
\caption{The Feynman diagrams for the background processes 
(a) $e^+e^-\to J/\psi \gamma$,
and (b) $e^+e^-\to e^+e^- J/\psi$.
\label{fig2}}
\end{figure*}

Charged-particle tracking is
provided by a five-layer silicon vertex tracker (SVT) and
a 40-layer drift chamber (DCH) operating in a 1.5-T axial
magnetic field. The transverse momentum resolution
is 0.47\% at 1~GeV/$c$. Energies of photons and electrons
are measured with a CsI(Tl) electromagnetic calorimeter
(EMC) with a resolution of 3\% at 1~GeV. Charged-particle
identification is provided by specific ionization 
measurements in the SVT and DCH, and by an internally reflecting
ring-imaging Cherenkov detector (DIRC). Electron identification
also uses shower shape in the EMC and the ratio of shower energy to track
momentum. Muons are identified in the solenoid's instrumented flux return,
which consist of iron plates interleaved with resistive plate chambers.

The signal process is simulated
with the Monte Carlo (MC) event generator GGResRc.  It uses the formula
for the differential cross section for pseudoscalar meson production
from Ref.~\cite{BKT}. Two samples of signal events are produced:
one for no-tag measurement without any kinematic restrictions, and the 
other with the restrictions on the momentum transfer values to the electrons 
$Q^2=-q_1^2 > 1.5$ GeV$^2$ and $-q_2^2 < 1$ GeV$^2$.
The restriction on $Q^2$ for the tagged electron corresponds to the detector 
acceptance. The experimental criteria providing these
restrictions for data events will be described in Sec.~\ref{evsel}.
In the simulation of no-tag events we use the form factor 
\begin{equation}
F(q_1^2,q_2^2)=\frac{F(0,0)}{(1-q_1^2/m_{J/\psi}^2)(1-q_2^2/m_{J/\psi}^2)}
\end{equation}
expected in the vector dominance model. The form factor is fixed to the 
constant value $F(0,0)$ in the simulation of single-tag events.
The produced $\eta_c$ decays into the $K_S K^\pm\pi^\mp$ final state. The 
simulation uses a three-body phase space distribution to describe this decay.   

The GGResRc event generator includes next-to-leading-order radiative
corrections to the Born cross section calculated according to Ref.~\cite{RC}.
In particular, it generates
extra photons emitted by
the initial- and final-state electrons.
The formulae from Ref.~\cite{RC} are modified to take into account
the hadron contribution to the vacuum polarization diagrams.
The maximum energy of the 
photon emitted from the initial state is restricted by the
requirement\footnote{Throughout this paper an asterisk superscript
denotes quantities in the $e^+e^-$ c.m. frame. In this frame the positive 
$z$-axis is defined to coincide with the $e^-$ beam direction.}
$E^\ast_\gamma < 0.05\sqrt{s}$, where $\sqrt{s}$ is the $e^+e^-$
c.m. energy.
The generated events are subjected to detailed
detector simulation based on GEANT4~\cite{GEANT4},
and are reconstructed with the
software chain used for the experimental data. Temporal variations in 
the detector performance
and beam background conditions are taken into account.

The processes with a $J/\psi$ in the final state (Fig.~\ref{fig2}),
with $J/\psi$ decaying into $\eta_c\gamma$, can imitate the process under study.
The initial state radiation (ISR) process (Fig.~\ref{fig2}(a)) contributes
to the no-tag mode, while the $J/\psi$ bremsstrahlung process 
(Fig.~\ref{fig2}(b)) contributes background to the single-tag mode.
We simulate both processes with $J/\psi$ decaying to $K_S K^\pm\pi^\mp$
and also to $\eta_c\gamma$ followed by 
$\eta_c\to K_S K^\pm \pi^\mp$. To estimate a possible
background from other two-photon processes we simulate the reaction 
$e^+e^-\to e^+e^-\eta_c\pi^0$. These events are generated with an 
isotropic distribution of the final state mesons in the $\eta_c\pi^0$
rest frame.

\section{Event selection\label{evsel}}
We select $e^+e^-\to e^+e^-\eta_c$ candidates in the 
no-tag and single-tag modes, with zero and one detected
electron, respectively. The decay mode 
$\eta_c \to K_S K^\pm\pi^\mp$ $(K_S \to \pi^+\pi^-)$ is used
to reconstruct $\eta_c$ meson candidates. 

Events are selected with at least four (five for single-tag mode) 
charged-particle tracks. Since a significant fraction of
events contain beam-generated spurious track and photon candidates, 
one extra track and any number of extra photons are allowed in
an event.
The tracks corresponding to the charged kaon and pion must be oppositely 
charged,
and must extrapolate to the interaction region. The kaon is required to be  
positively identified, while the pion track is required to be inconsistent 
with the kaon hypothesis.

The track identified as an electron must originate from the interaction region
and be in the polar angle range $0.387<\theta<2.400$ 
in the lab frame (0.64--2.69 in the $e^+e^-$ c.m. frame).
The latter requirement is needed for good electron identification. To 
recover electron energy loss due to bremsstrahlung, both internal and
in the detector material before the DCH, 
the energy of any EMC shower close to the electron direction is combined
with the measured energy of the electron track. The resulting c.m. energy of 
the electron candidate must be greater than 1 GeV.

A $K_S$ candidate is formed from a pair of oppositely charged tracks fitted 
to a common vertex, and yielding an invariant mass value in the range 
487.5--507.5 MeV/$c^2$, when the charged-pion mass is assigned to each track. 
The candidate is then refitted with a $K_S$ mass constraint to improve
the precision of its momentum measurement.
To suppress combinatorial background, the angle between the $K_S$ 
candidate momentum and the line
connecting its production and decay vertices ($\psi_{K_S}$) is required
to satisfy $\cos{\psi_{K_S}}>0.95$.

\begin{figure}
\includegraphics[width=.33\textwidth]{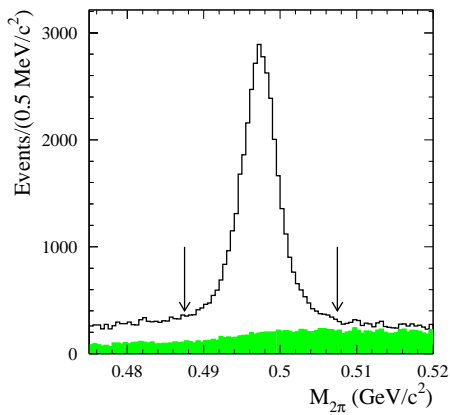}
\caption{The $M_{2\pi}$ distribution for $K_S$ candidates in the no-tag data
sample. The shaded histogram shows
events rejected by the requirement $\cos{\psi_{K_S}}>0.95$. The arrows
indicate the region used to select event candidates.
\label{fig3}}
\end{figure}

\begin{figure}
\includegraphics[width=.33\textwidth]{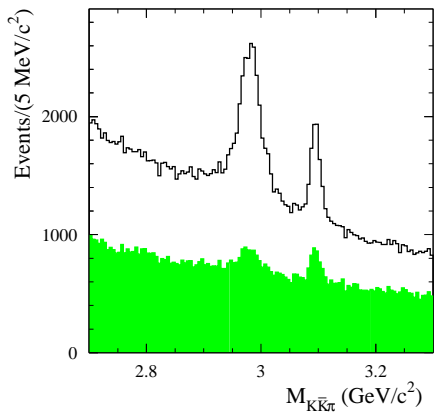}
\caption{The distribution of the invariant mass of $\eta_c$ candidates
in the no-tag data sample. The shaded histogram shows events rejected
by the requirement $p^\ast_\perp < 0.25$ GeV/$c$.
\label{fig4}}
\end{figure}

An $\eta_c$ candidate is formed from $K_S$, $K^\pm$, and $\pi^\mp$ candidates 
fitted to a common vertex. 
The preliminary selection criterion for no-tag events requires that
$|\cos{\theta^\ast_{\eta_c}}|>0.95$, where $\theta^\ast_{\eta_c}$ is the
polar angle of the candidate $\eta_c$ in the $e^+e^-$ c.m. frame. 
Figure~\ref{fig3} shows the distribution of the invariant mass of the pions 
forming a $K_S$ candidate ($M_{2\pi}$) for events satisfying this criterion.
The shaded histogram demonstrates the effect of the requirement 
$\cos{\psi_{K_S}}>0.95$. The transverse momentum of the $\eta_c$ candidate in
the $e^+e^-$ c.m. frame is restricted  to the range  $p^\ast_\perp < 0.25$ GeV/$c$. 
The invariant mass distribution for $\eta_c$ candidates is shown in 
Fig.~\ref{fig4}, where for events with more than one $\eta_c$ candidate
(about 0.4\% of signal events),
the candidate with the smallest value of $p^\ast_\perp$ is selected.   
The $\eta_c$ peak from two-photon production and the $J/\psi$ peak from 
the ISR process 
$e^+e^-\to J/\psi\gamma$ are clearly seen in the invariant mass distribution.
The shaded histogram shows the distribution for 
candidates rejected by the requirement $p^\ast_\perp < 0.25$ GeV/$c$.
This requirement limits the momentum transfers to
the electrons. The value of the effective threshold for
$q^2_i$ $(i=1,2)$
is determined from the dependence of the detection efficiency on
$\max(-q^2_1,-q^2_2)$ and is about 0.1 GeV$^2$. Such a low $q^2$ threshold 
yields a model-independent extraction of $F(0,0)$ from the no-tag data.
We note that ${\rm d}\sigma/{\rm d}q_1^2{\rm d}q_2^2\propto 1/(q^2_1q^2_2)$ 
at small $|q^2_1|$ and $|q^2_2|$.

\begin{figure}
\includegraphics[width=.33\textwidth]{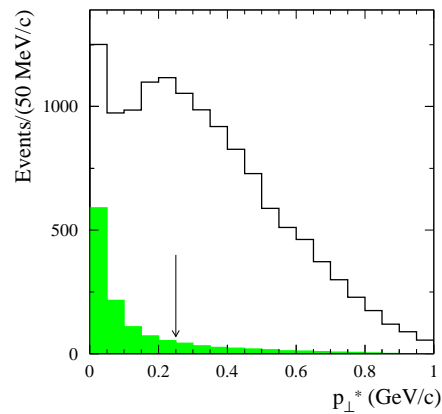}
\caption{The transverse momentum distribution for $e \eta_c$ data candidates.
The shaded histogram is for simulated signal events. Data candidates for which
$p^\ast_\perp < 0.25$ GeV/$c$ (indicated by the arrow) are retained.
\label{fig5}}
\end{figure}

For single-tag events we combine an $\eta_c$ candidate with
an electron candidate and
require $|\cos{\theta^\ast_{e\eta_c}}|>0.95$, where $\theta^\ast_{e\eta_c}$
is the polar angle of the momentum vector of the $e\eta_c$ system
in the $e^+e^-$ c.m. frame. 
The transverse momentum of the
$e \eta_c$ system is restricted to  $p^\ast_\perp < 0.25$ GeV/$c$. The 
$p^\ast_\perp$ distribution for data candidates is shown in Fig.~\ref{fig5},
where 
the shaded histogram is the corresponding distribution for simulated signal
events. The condition on $p^\ast_\perp$ limits the value of the momentum
transfer to the untagged electron ($q_2^2$). The effective $q^2_2$ threshold
determined from simulation is about 0.1 GeV$^2$.
\begin{figure}
\includegraphics[width=.33\textwidth]{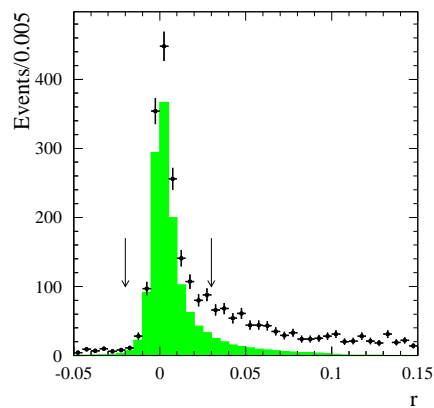}
\caption{The  distribution of $r$ defined in Eq.(\ref{eq_r}) 
for data (points with error bars) and simulated signal events
(shaded histogram). The arrows indicate the region used to select
candidate events ($-0.02<r<0.03$).
\label{fig6}}
\end{figure}

The emission of extra photons from the electrons involved 
leads to a difference between
the measured and actual values of $Q^2$. In the case of ISR,
$Q^2_{\rm meas}=Q^2_{\rm true}(1+r_\gamma)$, where 
$r_\gamma=2E^\ast_\gamma/\sqrt{s}$.
To restrict the energy of the ISR photon we use the parameter
\begin{equation}
r=\frac{{\sqrt{s}}-E^\ast_{e\eta_c}-|p^\ast_{e\eta_c}|}{\sqrt{s}},
\label{eq_r}
\end{equation}
where $E^\ast_{e\eta_c}$ and $p^\ast_{e\eta_c}$ are the c.m. energy and 
momentum of the detected $e\eta_c$ system. In the ISR case this parameter   
coincides with $r_\gamma$ defined above. The $r$ distributions for data and
simulated signal are shown in Fig.~\ref{fig6}. Candidates with $-0.02<r<0.03$ 
are retained. We note that the condition on $r$ ensures
compliance with the restriction $r_{\gamma}<0.1$ used in the MC simulation.

For two-photon events with a tagged positron (electron),
the momentum of the detected $e\eta_c$ system in the $e^+e^-$ c.m.
frame has a negative (positive) $z$-component, while events resulting 
from $e^+e^-$ annihilation are produced symmetrically.
To suppress the $e^+e^-$ annihilation background, event candidates with the 
wrong sign of the  momentum  $z$-component are removed.

\begin{figure}
\includegraphics[width=.33\textwidth]{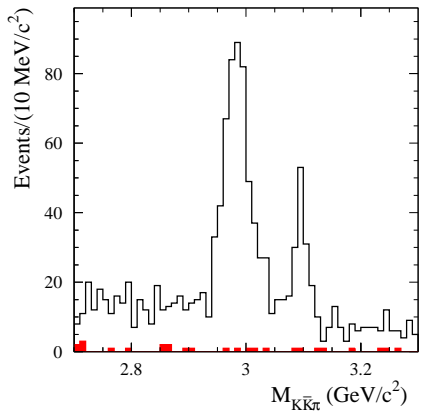}
\caption{The invariant mass distribution for single-tag $\eta_c$ candidates.
The shaded histogram is for events 
with the wrong sign of the $e\eta_c$ momentum  $z$-component.
\label{fig7}}
\end{figure}
The distribution of the invariant mass
of $\eta_c$ candidates for single-tag events satisfying the selection
criteria described above is shown in Fig.~\ref{fig7}. For events with more 
than one $\eta_c$ candidate, the candidate with smallest $p^\ast_\perp$ is 
selected. 
Signals corresponding to $\eta_c$ and $J/\psi$ production are seen clearly 
in the mass spectrum. The $J/\psi$ events are from the process 
$e^+e^-\to e^+e^-J/\psi$ (see Fig.~\ref{fig2}(b)). The shaded histogram in 
Fig.~\ref{fig7}
shows the distribution for candidates with the wrong sign of the $e\eta_c$ 
momentum  $z$-component. Since the numbers of events from
$e^+e^-$ annihilation in the wrong- and right-sign data samples are
expected to be approximately the same, this shows that the background   
from $e^+e^-$ annihilation peaking at the $\eta_c$ mass is small.
 
\section{Fitting the $K_SK^\pm\pi^\mp$  mass spectrum for no-tag events}\label{fitting}
\begin{figure}
\includegraphics[width=.33\textwidth]{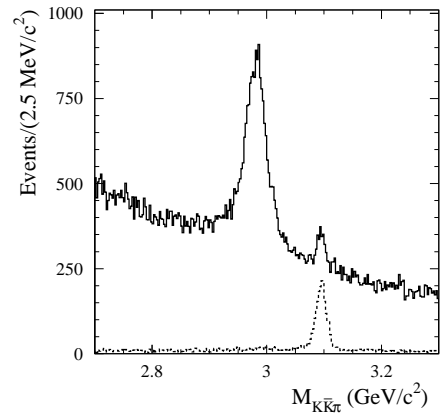}
\caption{The $K_SK^\pm\pi^\mp$ invariant mass spectrum for
no-tag data events satisfying (dashed histogram) and not satisfying
(solid histogram) the condition for ISR event selection
(Eq.(\ref{ISRsel})).
\label{fig8}}
\end{figure}

The mass spectrum for no-tag events exhibits the $\eta_c$ and
$J/\psi$ peaks corresponding to the two-photon and ISR
production mechanisms, respectively. 
The c.m. momentum $p^\ast$ of the $K_S K^\pm\pi^\mp$ system for ISR events
is equal to $(\sqrt{s}/2)(1-M_{K\bar{K}\pi}^2/s)$. In the mass region
under study the detector acceptance strongly limits the efficiency for ISR 
events. Due to the asymmetry of the acceptance
most of the detected ISR events have positive 
$\cos{\theta^\ast_{\eta_c}}$. It follows that the ISR events can be selected 
by requiring:
\begin{equation}
p^\ast/(1-M_{K\bar{K}\pi}^2/s) > 5.1\mbox{ GeV/$c$, } 
 \cos{\theta^\ast_{\eta_c}} < 0.\label{ISRsel} 
\end{equation} 
The mass distribution for the events satisfying this condition is shown by 
dashed histogram in Fig.~\ref{fig8}. The selected event sample contains 
mostly ISR events with very little two-photon $\eta_c$ admixture.    
The $K_S K^\pm\pi^\mp$ mass distribution for the events not satisfying
Eq.(\ref{ISRsel}) is shown by the solid histogram in  Fig.~\ref{fig8}.
The remaining $J/\psi$ events are  
from the ISR process with more than one photon emitted from the initial state.  

To determine the $\eta_c$ mass and width, and the 
number of events containing an $\eta_c$, a binned likelihood 
fit is made to the distributions in Fig.~\ref{fig8} using
a function consisting of signal ($\eta_c$ and $J/\psi$) and background
distributions. The bin size used in the fit is chosen to be 2.5 MeV/$c^2$.
The $J/\psi$ line shape is represented by the detector 
resolution function for ISR events. The $\eta_c$ line shape is described by a 
Breit-Wigner function convolved with the detector resolution function 
corresponding to two-photon production. In each case,
the detector resolution function is obtained using MC simulation of the 
detector response.
We use the nonrelativistic Breit-Wigner form
\begin{equation}
\frac{(\Gamma/2)^2}{(M_0-M)^2+(\Gamma/2)^2},
\end{equation}
where $M$ is the $K_SK^\pm\pi^\mp$ invariant mass, and $M_0$ and 
$\Gamma$ are the $\eta_c$ mass and width. 
The changes in the values of the parameters, if a relativistic Breit-Wigner
function is used, are negligible.

\begin{figure*}
\includegraphics[width=.4\textwidth]{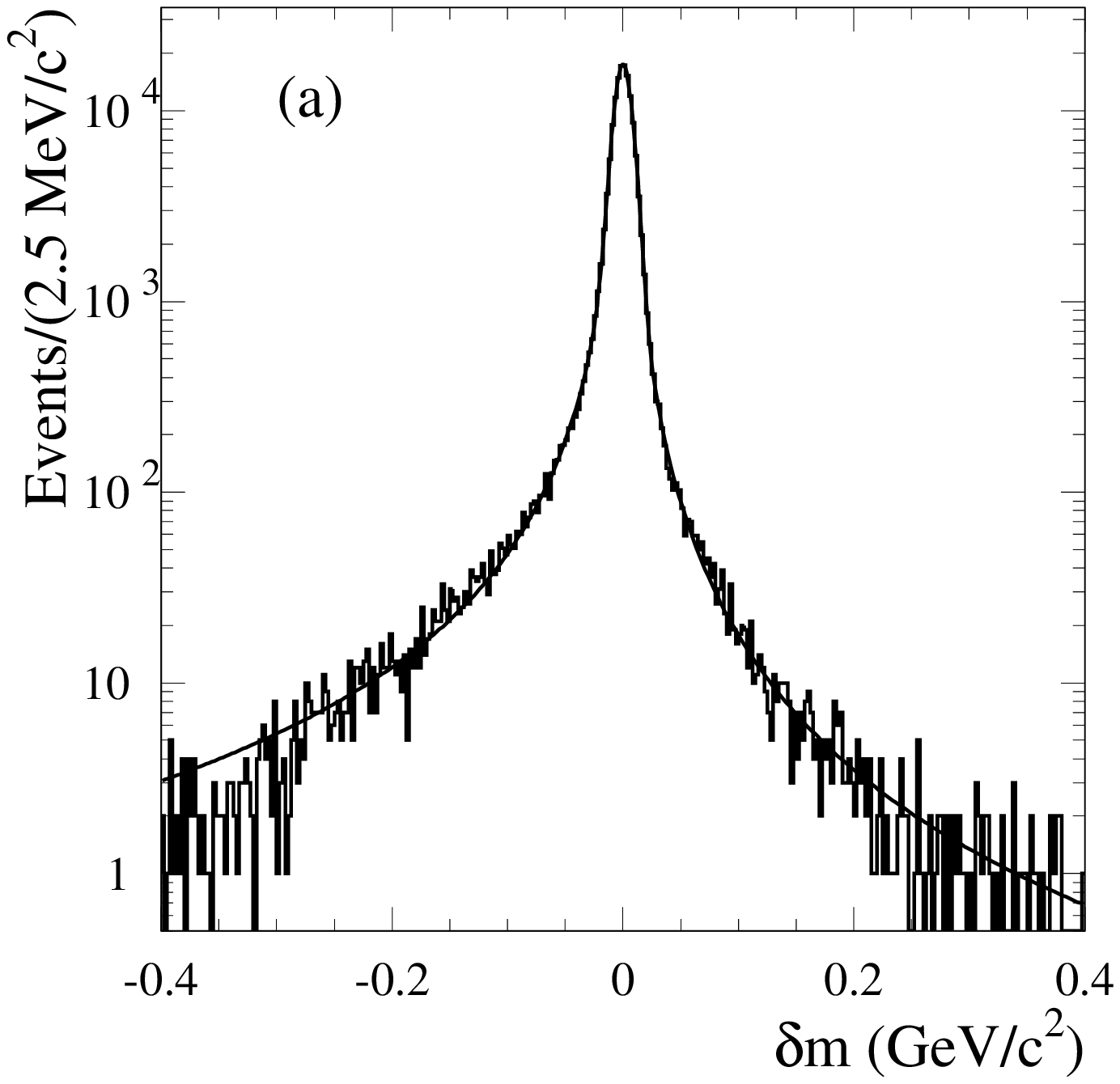}
\hfill
\includegraphics[width=.4\textwidth]{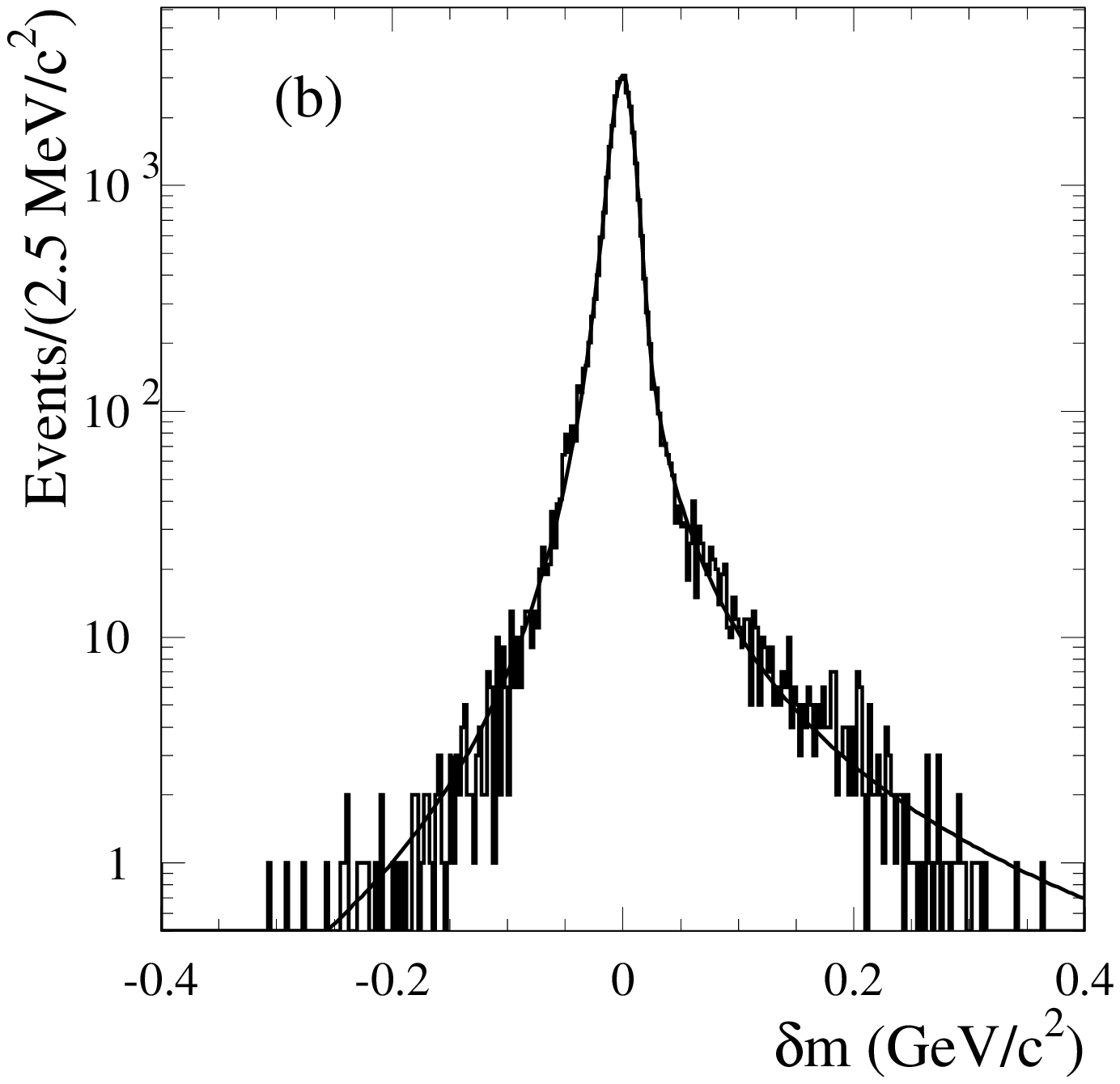}
\caption{The distribution of the difference ($\delta m$) between measured and 
true $K_SK^\pm\pi^\mp$ mass for simulated events (a) for the two-photon
process $e^+e^-\to e^+e^-\eta_c$, and (b) for the ISR process
$e^+e^-\to J/\psi \gamma$. The curves correspond to fits defined in
the text.\label{fig9}}
\end{figure*}

The detector resolution functions are determined from
the distributions of the difference between measured and true 
simulated $K_SK^\pm\pi^\mp$ mass for the processes 
$e^+e^-\to e^+e^-\eta_c,\;\eta_c\to K_SK^\pm\pi^\mp$
and $e^+e^-\to J/\psi \gamma,\;J/\psi\to K_SK^\pm\pi^\mp $ 
shown in Fig.~\ref{fig9}(a) and Fig.~\ref{fig9}(b), respectively.
The distributions are fit with the following function:
\begin{equation}
F(x)=A[G(x)\sin^2{\zeta}+B(x)\cos^2{\zeta}],
\label{func}
\end{equation} 
where
\begin{equation}
G(x)=\exp \left (-\frac{(x-a_1)^2}{2\sigma^2} \right ),
\end{equation}
\begin{equation}
B(x)=\left\{ \begin{array}{ll}
\frac{(\Gamma_1/2)^{\beta_1}}{(a_2-x)^{\beta_1}+(\Gamma_1/2)^{\beta_1}} &
\mbox{if $x < a_2$};\\
\frac{(\Gamma_2/2)^{\beta_2}}{(x-a_2)^{\beta_2}+(\Gamma_2/2)^{\beta_2}} &
\mbox{if $x \geq a_2$},
\end{array}\right.
\end{equation}
$A$, $\zeta$, $a_1$, $\sigma$, $a_2$, $\Gamma_1$, $\beta_1$,
$\Gamma_2$, $\beta_2$ are free fit parameters.
The $B(x)$ term is added to the Gaussian function to describe the asymmetric 
power-law tails of the $\delta m$ distributions.

When used in data, the resolution $\sigma$ in the Gaussian term of 
Eq.(\ref{func}), is modified to take into account a possible  
difference between data and simulation:
\begin{equation}
\sigma^2=\left\{ \begin{array}{ll}
\sigma^2_{\rm MC}-\Delta\sigma^2 & \mbox{if $\Delta\sigma < 0$};\\
\sigma^2_{\rm MC}+\Delta\sigma^2 & \mbox{if $\Delta\sigma \geq 0$}.
\end{array}\right.
\end{equation}
The parameter  $\sigma_{\rm MC}$ is found to be 7.8 MeV/$c^2$ for the $J/\psi$ 
and $7.6$ MeV/$c^2$ for the $\eta_c$.
The parameter $\Delta\sigma$ is determined from the fit to the 
measured $K_SK^\pm\pi^\mp$ mass spectra.

\begin{figure*}
\includegraphics[width=.56\textwidth]{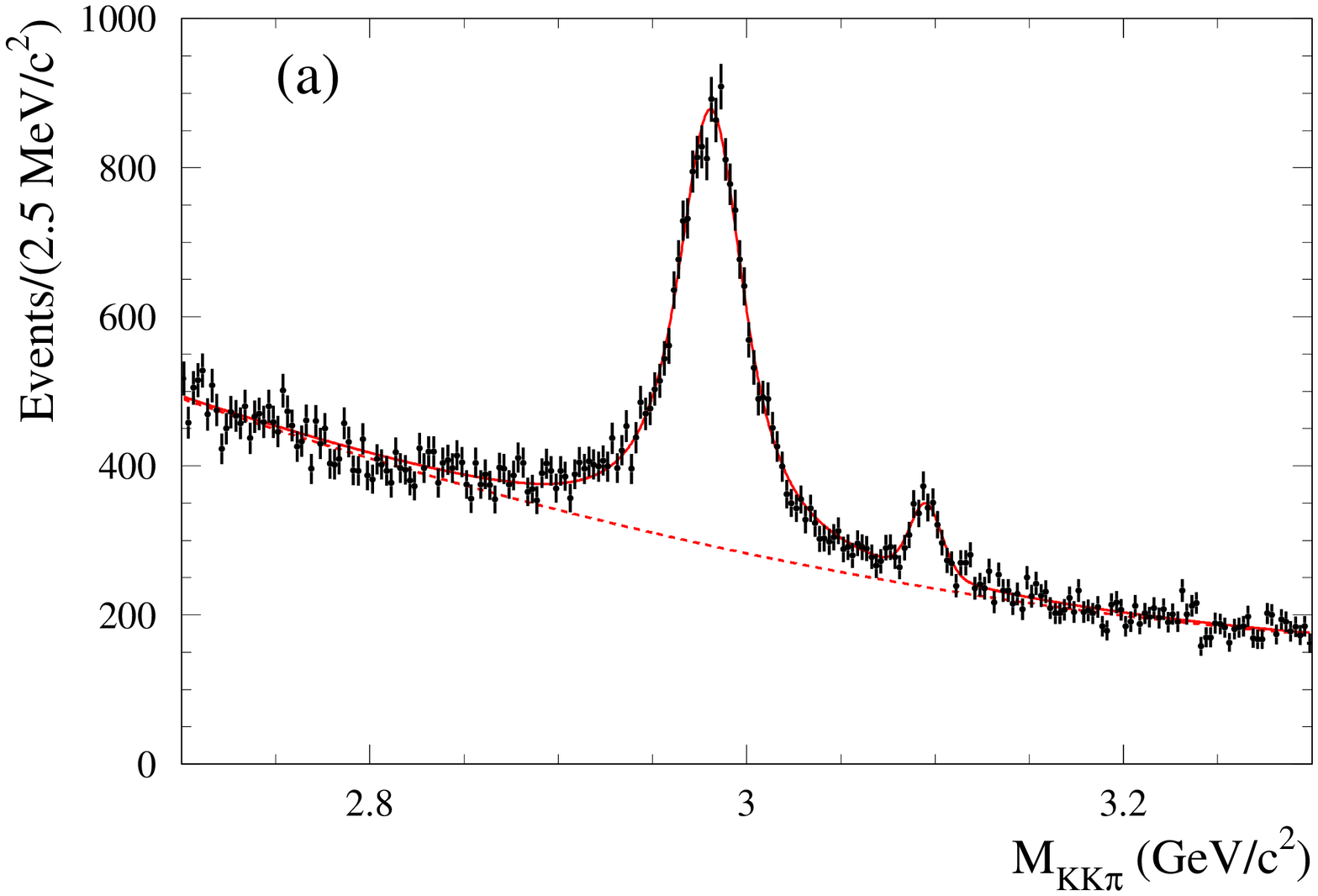}
\hfill
\includegraphics[width=.4\textwidth]{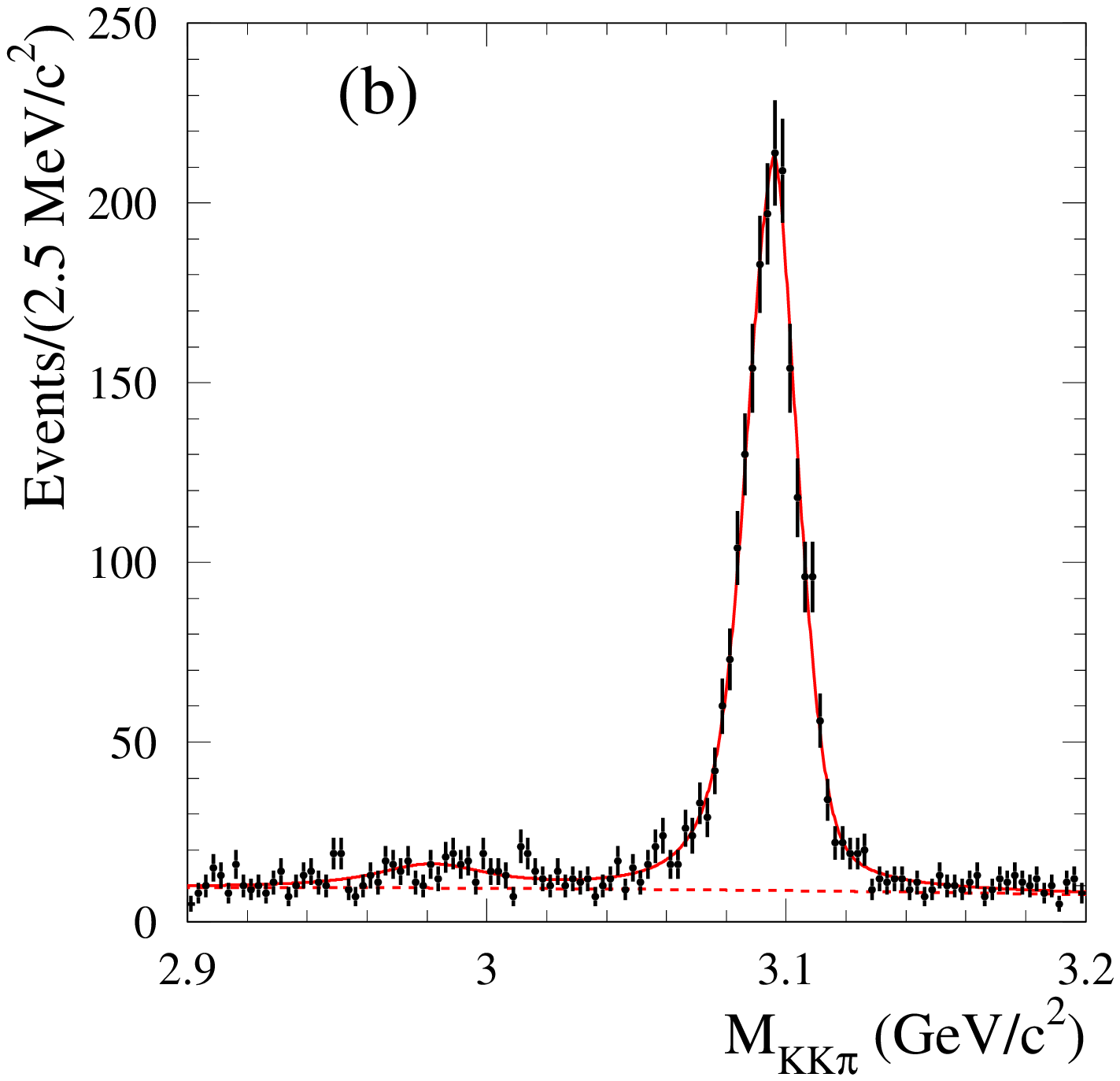}
\caption{The $K_SK^\pm\pi^\mp$ invariant mass distribution and fitted
curve for no-tag data events that (a) fail Eq.(\ref{ISRsel}), and (b) 
satisfy Eq.(\ref{ISRsel}) for ISR event selection.
\label{fig10}}
\end{figure*}

The background distribution is described by a second-order polynomial.
Both spectra shown in Fig.~\ref{fig8} are fit simultaneously
with 14 free parameters: the $J/\psi$ peak position ($m_{J/\psi}$), 
the difference between the $J/\psi$ and $\eta_c$ mass values ($\Delta m$), 
the $\eta_c$ width ($\Gamma$), 
the numbers of $\eta_c$ and $J/\psi$ events, $\Delta\sigma$, 
and the background parameters for both spectra. 
The fitted curves are shown in Fig.~\ref{fig10}. 
For the full mass range, 2.7--3.3 GeV/$c^2$, the $\chi^2$ values 
corresponding to the $\eta_c$ and $J/\psi$ distributions (solid and dashed 
in Fig.~\ref{fig8}) are 230 and 198, respectively, for a total number of 
degrees of freedom $2\times240-14$. 
The resulting parameter values are as follows:
\begin{eqnarray}
\Delta m &=& 114.7\pm0.4 \mbox{ MeV/$c^2$},\nonumber\\
\Gamma &=& 31.7\pm1.2 \mbox{ MeV},\nonumber\\
N_{\eta_c} &=& 14450\pm320,\nonumber\\
\Delta\sigma &=& -0.4\pm2.5 \mbox{ MeV/$c^2$},\nonumber\\
m_{J/\psi} &=& 3095.8\pm0.3 \mbox{ MeV/$c^2$}.
\end{eqnarray}
The mass resolution for the $J/\psi$ in data is
found to be consistent with the prediction of MC simulation.
The fitted value of the $J/\psi$ mass is shifted by $-1.1\pm 0.3$ MeV/$c^2$
from the nominal $J/\psi$ mass value~\cite{pdg}. It is assumed that this mass 
scale shift does not affect the mass difference $\Delta m$.
Since the momentum distributions for $J/\psi$ and $\eta_c$ events 
are different, and the MC simulation of the detector response is
not perfect, we test this assumption as follows.
The no-tag event sample was divided into three subsamples with approximately 
equal statistics
but with different laboratory $z$ momentum ($p_z$) of $\eta_c$ candidates. 
The average 
$p_z$ values in the subsamples are 3.2, $-0.4$, and $-1.3$ GeV/$c$, 
while the $J/\psi$ momentum is peaked at $p_z=-2.34$ GeV/$c$. The fitted 
values of the $\Delta m$ parameter
for the three subsamples are found to be shifted relative to the nominal
fit value by $0.5\pm0.6$, $-0.6\pm0.6$, and $0.2\pm0.6$ MeV/$c^2$, 
respectively. We do not observe any significant dependence of the
$\Delta m$ parameter on the $\eta_c$ momentum direction and absolute
value. Nevertheless, the shift value at the maximum difference between
the $\eta_c$ and $J/\psi$ momenta,  
0.5 MeV/$c^2$, is taken as an estimate of the $\Delta m$ systematic 
uncertainty due to the difference of the $J/\psi$ and $\eta_c$ momentum 
distributions.

To estimate the uncertainty of the fit parameters due to the assumed 
background shape, the second-order polynomial describing background 
in Fig.~\ref{fig10}(a) is replaced by an exponential function, and the
changes in the parameter values are considered to be measures of 
their associated systematic uncertainties.
This yields $\Delta\Gamma=0.8$ MeV and $\Delta N_{\eta_c}=400$. 

\begin{figure*}
\includegraphics[width=.4\textwidth]{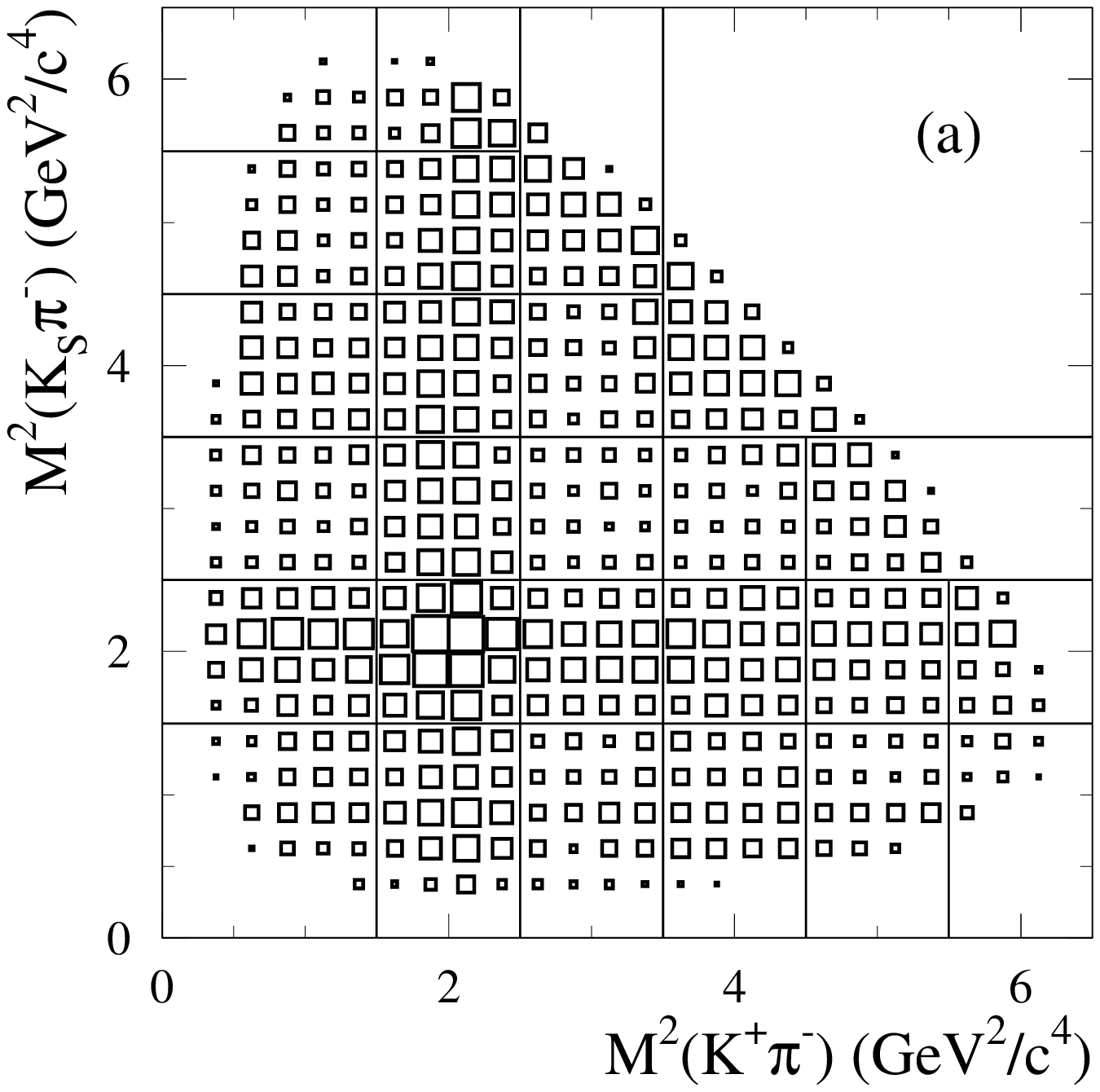}
\hfill
\includegraphics[width=.4\textwidth]{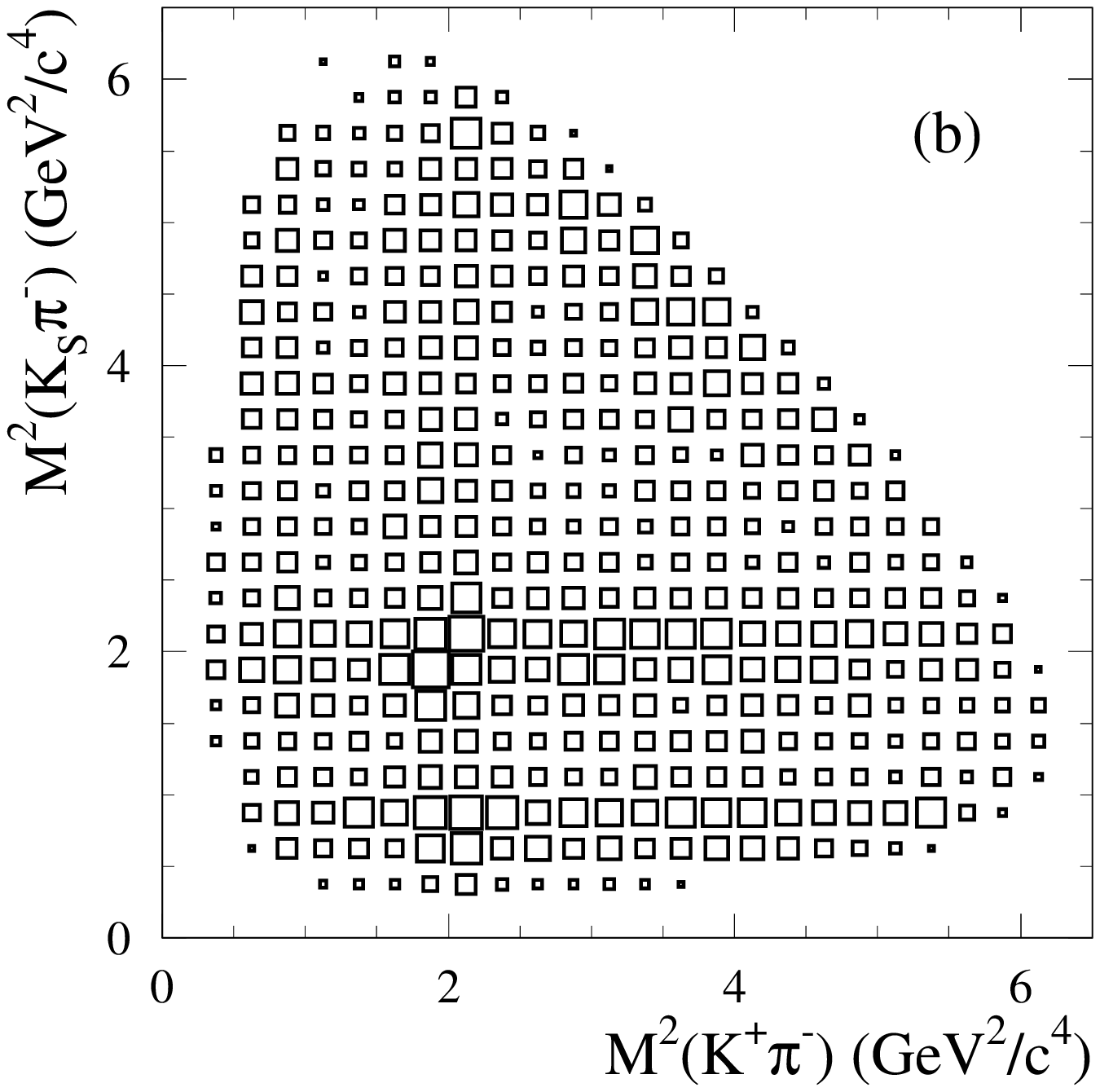}
\caption{The Dalitz plots for data events from  (a) the $\eta_c$ signal region
$2.94<M_{K\bar{K}\pi}<3.02$ GeV/$c^2$, and (b) the combined sideband regions,
$2.90<M_{K\bar{K}\pi}<2.94$ GeV/$c^2$ and $3.02<M_{K\bar{K}\pi}<3.06$ GeV/$c^2$.
\label{fig11}}
\end{figure*}

\begin{figure*}
\includegraphics[width=.32\textwidth]{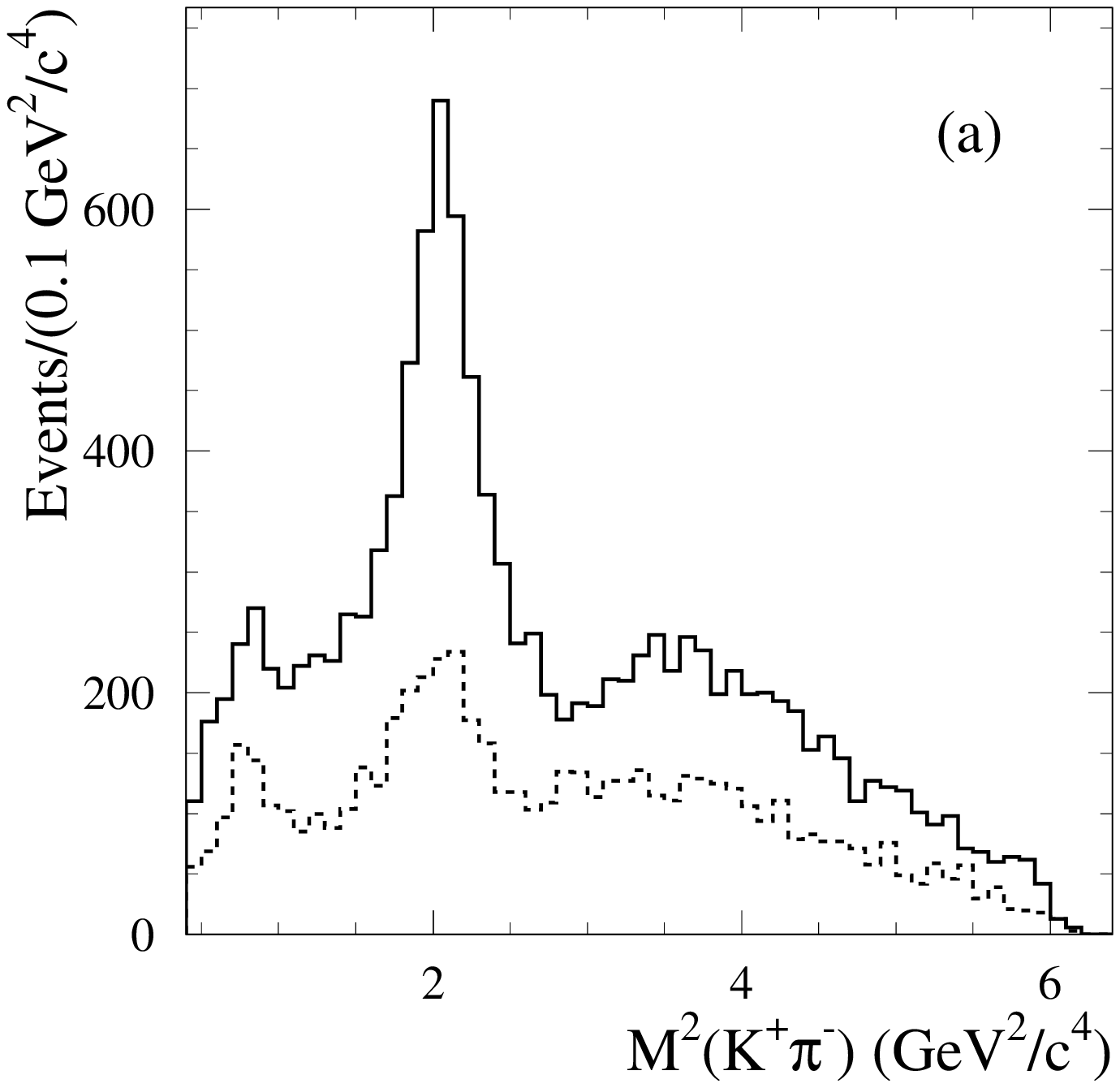}
\includegraphics[width=.32\textwidth]{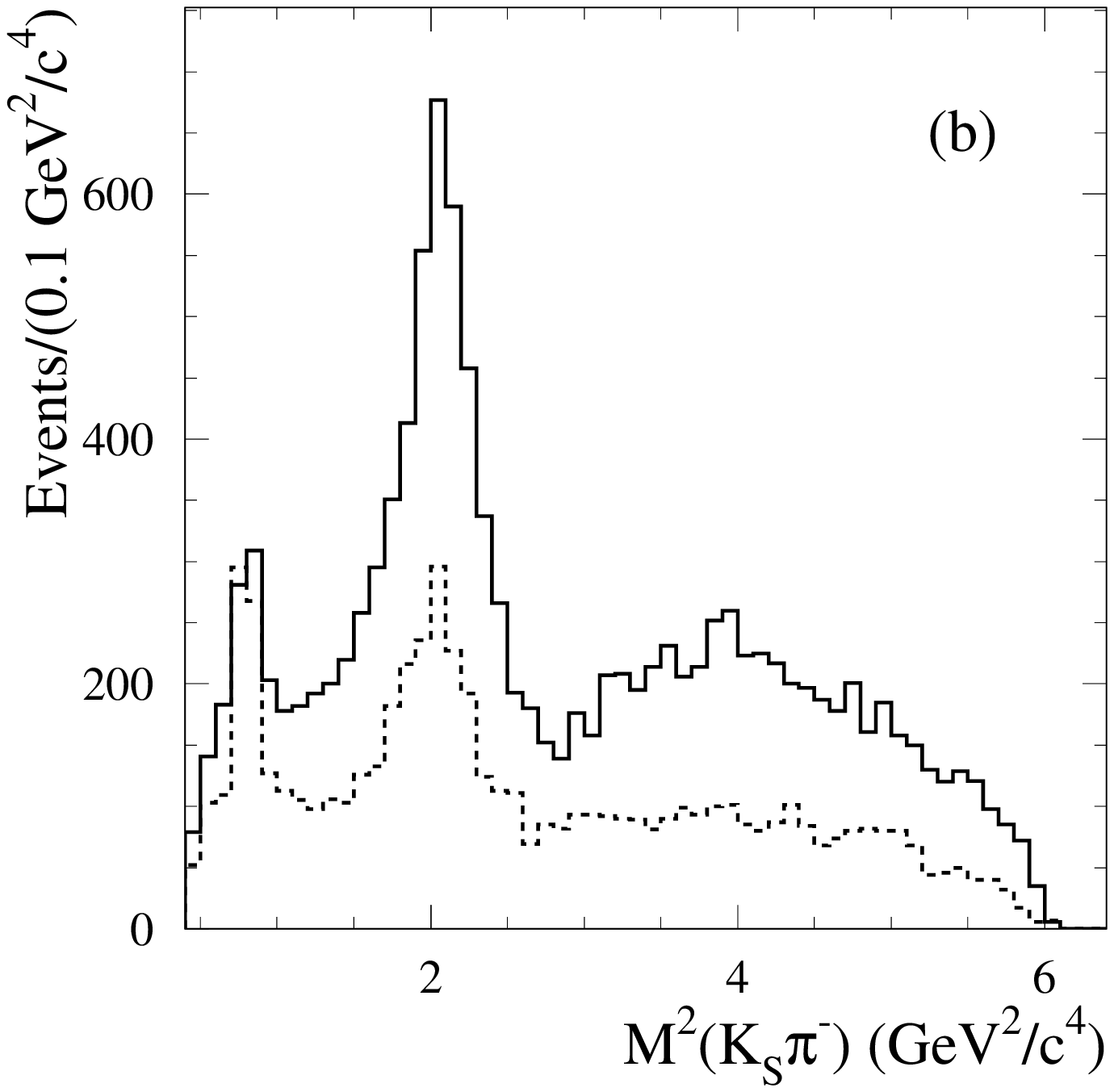}
\hfill
\includegraphics[width=.32\textwidth]{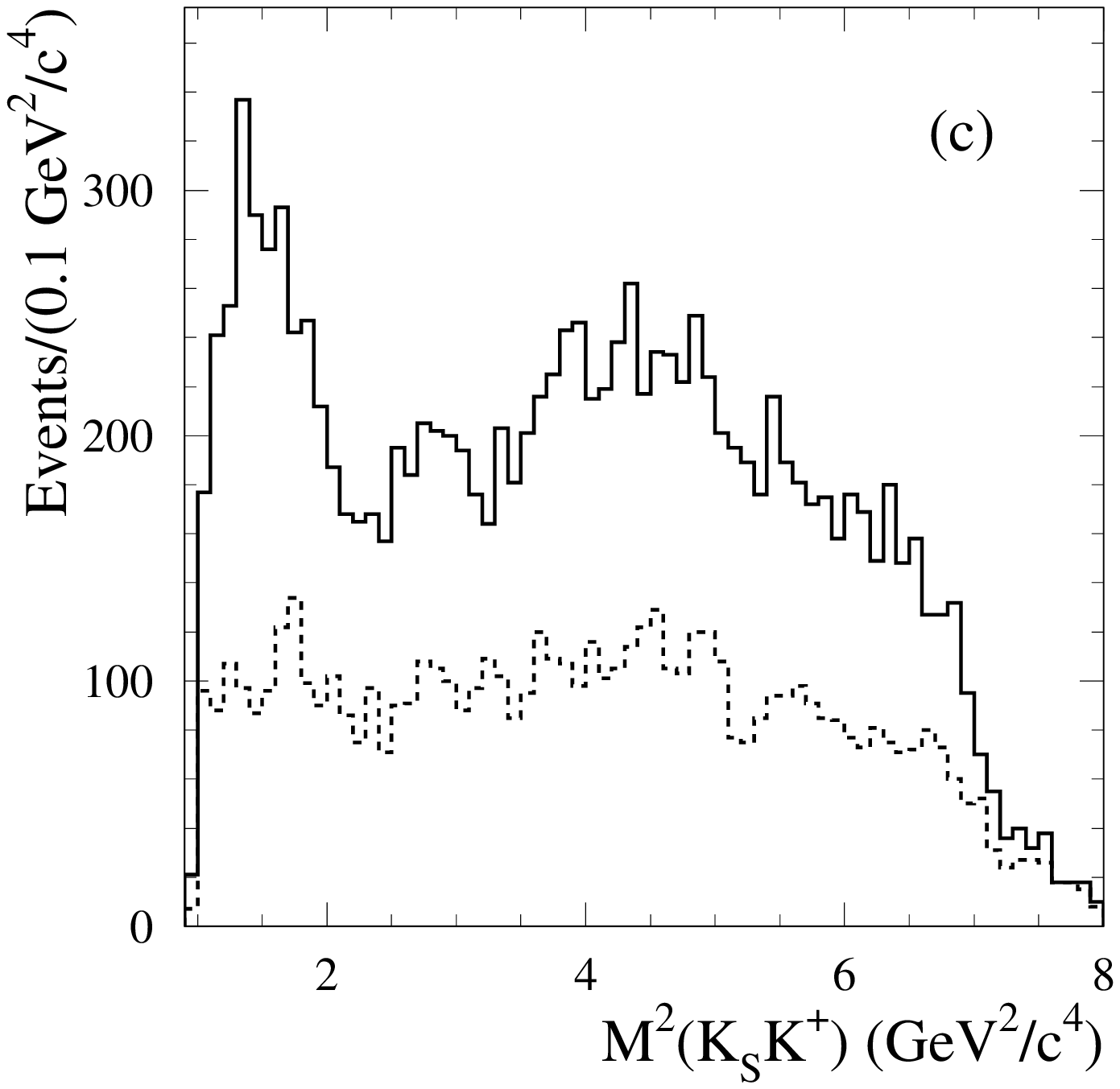}
\caption{The projections of the Dalitz plots of Fig.~\ref{fig11};
the solid and dashed histograms correspond to the $\eta_c$ signal and sideband 
$M_{K\bar{K}\pi}$ regions, respectively.
\label{fig12}}
\end{figure*}

The MC simulation uses a phase space distribution for 
$\eta_c\to K_S K^\pm\pi^\mp$ decay. This can lead to distortion of the 
resolution function and a systematic change in the detection efficiency 
determined from simulation. In order to address this issue, a study of the 
Dalitz plot distribution for 
$\eta_c\to K_S K^\pm\pi^\mp$ decay was performed. 
The Dalitz plots for data events from the
$\eta_c$ signal ($2.94<M_{K\bar{K}\pi}<3.02$ GeV/$c^2$) and sideband 
($2.90<M_{K\bar{K}\pi}<2.94$ GeV/$c^2$ and $3.02<M_{K\bar{K}\pi}<3.06$ 
GeV/$c^2$) regions are
shown in Fig.~\ref{fig11}, and their projections are shown
in Fig.~\ref{fig12}. 

\begin{figure}
\includegraphics[width=.4\textwidth]{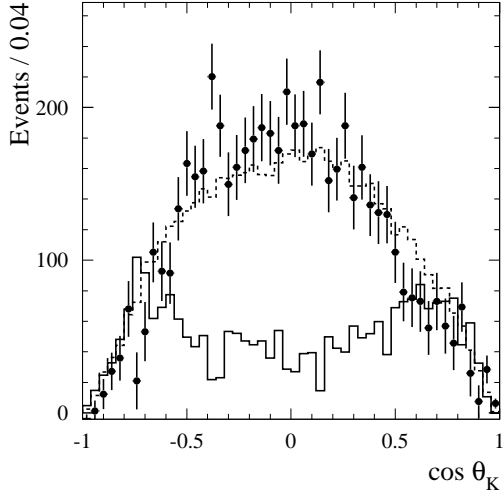}
\caption{The $\cos{\theta_K}$ distribution for events
from the $\eta_c$ signal region after background subtraction (points with error
bars), and for events from the sidebands after subtraction of the $\eta_c$ 
contribution (solid histogram). The dashed histogram represents the signal
MC simulation.
\label{fig13}}
\end{figure}

Figure~\ref{fig13} shows the distributions of $\cos{\theta_K}$, where 
$\theta_K$ is the charged-kaon polar angle in the $K_S K^\pm\pi^\mp$ rest
frame, for events from the $\eta_c$ region after background subtraction,
and for events from the sidebands with the $\eta_c$ contribution subtracted.
The $\cos{\theta_K}$ distribution for $\eta_c$ events closely resembles
the distribution in the signal MC simulation shown by the dashed histogram
in Fig.\ref{fig13}.

Since the MC simulation uses
a phase space decay model, this suggests that in the $K_S K^\pm\pi^\mp$ mode 
the $\eta_c$ decays predominantly via the scalar $K^\ast_0(1430)$ meson, i.e.,
$\eta_c\to K^\ast_0(1430)\bar{K}$. 

It should be noted that a significant part of the 
non-$\eta_c$ background comes from $\gamma\gamma\to K_S K^\pm\pi^\mp$, 
and so can interfere with the signal process
$\gamma\gamma\to\eta_c\to K_S K^\pm\pi^\mp$. From the Dalitz plot,
the background process seems to proceed mainly via the intermediate
$K^\ast(892)\bar{K}$ and $K^\ast(1430)\bar{K}$ states, where the $K^\ast(1430)$ 
may be either the scalar ($K^\ast_0(1430)$) or tensor ($K^\ast_2(1430)$) state.
Interference between the $I=0$ and $I=1$ background amplitudes 
appears to lead to the suppression of neutral $K^\ast(892)$ 
production (observed in Fig.~\ref{fig12}(a) with respect to
Fig.~\ref{fig12}(b)). 
The $\eta_c$ signal distribution in Fig.~\ref{fig13} 
actually represents the $\cos{\theta_K}$ dependence of the detector acceptance.
The $\cos{\theta_K}$ distribution for non-$\eta_c$ 
events corrected for this acceptance will be peaked near $\pm 1$.
This means that the kaon in continuum is dominantly produced with
large orbital momentum.
Therefore, interference between the $S$-wave $\eta_c$ decay amplitude and
the $\gamma\gamma\to K_S K^\pm\pi^\mp$ nonresonant amplitude is expected to 
be small.

To estimate possible shifts of the $\eta_c$ parameter values due to 
interference, 
an interference term is introduced into the fitting function
through the following:
\begin{equation}
\left | \frac{\Gamma/2}{M_0-M-i\Gamma/2}+
A\sqrt{\frac{P_2(M)}{P_2(M_0)}}\right |^2,
\end{equation}
where the Breit-Wigner function describes the $\eta_c$ amplitude, $P_2(M)$ is a
second-order polynomial describing the mass dependence of the nonresonant 
intensity, and $A$ is the value of this amplitude at the $\eta_c$ mass.
The $P_2(M)$ coefficients are chosen to be equal to the coefficients of
the second order polynomial describing the nonresonant background.
A comparison of the Dalitz plot distributions for $\eta_c$ and non-$\eta_c$
events indicates that the maximal interference should occur in the
vicinity of $M^2(K\pi)\approx2$ GeV$^2/c^4$, where in both signal and 
background the quasi-two-body final states, $K^\ast_{0}(1430)\bar{K}$ and 
$K^\ast_2(1430)\bar{K}$, dominate. Therefore, no significant
phase shift is expected between the two amplitudes, and so parameter $A$
is chosen to be real. From the fit, $A=0.03\pm0.01$ and there are 
insignificant 
changes in the values of parameters $\Gamma$ and $N_{\eta_c}$. 
The value of $\Delta m$ changes by 1.5 MeV/$c^2$, and so this is considered to
provide an estimate of the systematic uncertainty due 
to possible interference between the $\eta_c$ and nonresonant amplitudes. 

To take into account the difference between data and simulation of
the $\eta_c\to K_S K^\pm\pi^\mp$ decay dynamics,
the Dalitz plot is divided into 26 cells as shown in Fig.~\ref{fig11}(a).
For each cell the $K_SK^\pm\pi^\mp$ mass spectrum is fit using
the fitting function described above, and the number of $\eta_c$
events is determined. This experimental Dalitz plot distribution corrected 
for detection efficiency is used to reweight the signal simulation.
The reweighting changes the resolution function and the fit parameter values 
insignificantly. 

Thus, from the fit to the $K_SK^\pm\pi^\mp$ mass spectrum for no-tag
events the following $\eta_c$ parameter values are obtained:
\begin{eqnarray}
\Delta m &=& 114.7\pm0.4\pm1.6 \mbox{
MeV/$c^2$},\nonumber\\
\Gamma &=& 31.7\pm1.2\pm0.8 \mbox{ MeV},\nonumber\\
N_{\eta_c} &=& 14450\pm320\pm400.
\end{eqnarray}
When the nominal value of the $J/\psi$ mass~\cite{pdg} is used,
the $\eta_c$ mass becomes $2982.2\pm0.4\pm1.6$ MeV/$c^2$. 
The results for the mass and
width are in agreement with the previous \babar\ measurement obtained using
88 fb$^{-1}$ data~\cite{bb_etac}:
$m_{\eta_c}=2982.5\pm1.1\pm0.9$ MeV/$c^2$ and $\Gamma=34.3\pm2.3\pm0.9$
MeV. The systematic uncertainty reported here is greater than that reported
in the previous analysis as we are now allowing for the possibility that
there exists a $J^P=0^-$ continuum $K_SK^\pm\pi^\mp$ amplitude which 
interferes with the $\eta_c$ amplitude. 

\section{Fitting the $K_SK^\pm\pi^\mp$  mass spectrum for single-tag events}
\begin{figure}
\includegraphics[width=.4\textwidth]{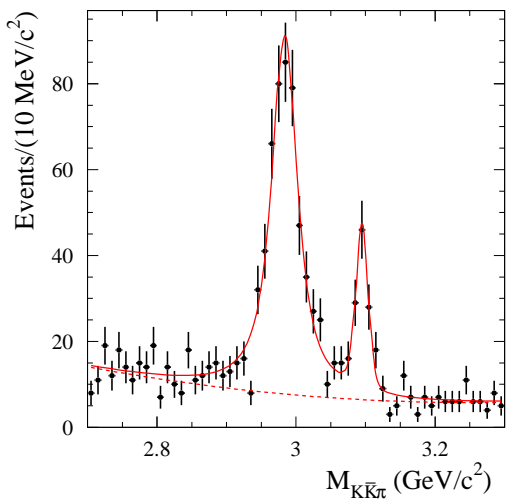}
\caption{The $K_SK^\pm\pi^\mp$ mass spectrum for single-tag
data events with $2 < Q^2  <50$ GeV$^2$. The solid curve is the
fit result. The dashed curve represents non-peaking background.
\label{fig14}}
\end{figure}

The $K_SK^\pm\pi^\mp$ mass spectrum for single-tag events from data
with $2 < Q^2  <50$ GeV$^2$ is shown in Fig.~\ref{fig14}. For $Q^2 > 50$
GeV$^2$ we do not see evidence of an $\eta_c$ signal over background.
To determine the number of $\eta_c$ events, a binned likelihood
fit is performed to the spectrum using a function consisting of a sum of
$\eta_c$, $J/\psi$, and background distributions. 
The bin size used in the fit is chosen to be 2.5 MeV/$c^2$.
The mass resolution line shape is described by the function of
Eq.~(\ref{func}) with parameters  determined from the signal simulation
reweighted to reproduce the $Q^2$ dependence observed in data.
This resolution function and its convolution with a Breit-Wigner function
are used to describe the $J/\psi$ and $\eta_c$ line shapes, respectively.
The background distribution is described by either a second order polynomial
or an exponential function. The fit result is shown in Fig.~\ref{fig14} for
a quadratic background. The fitted $\eta_c$ parameter values, 
$\Delta m=111.2\pm2.0$ MeV/$c^2$ and $\Gamma=31.9\pm4.3$ MeV,
are in agreement with the results obtained for no-tag events, and
the total $\eta_c$ signal is $530\pm41$ events. The difference
in signal yield for the two background hypotheses is 17 events. 

\begin{figure*}
\includegraphics[width=.32\textwidth]{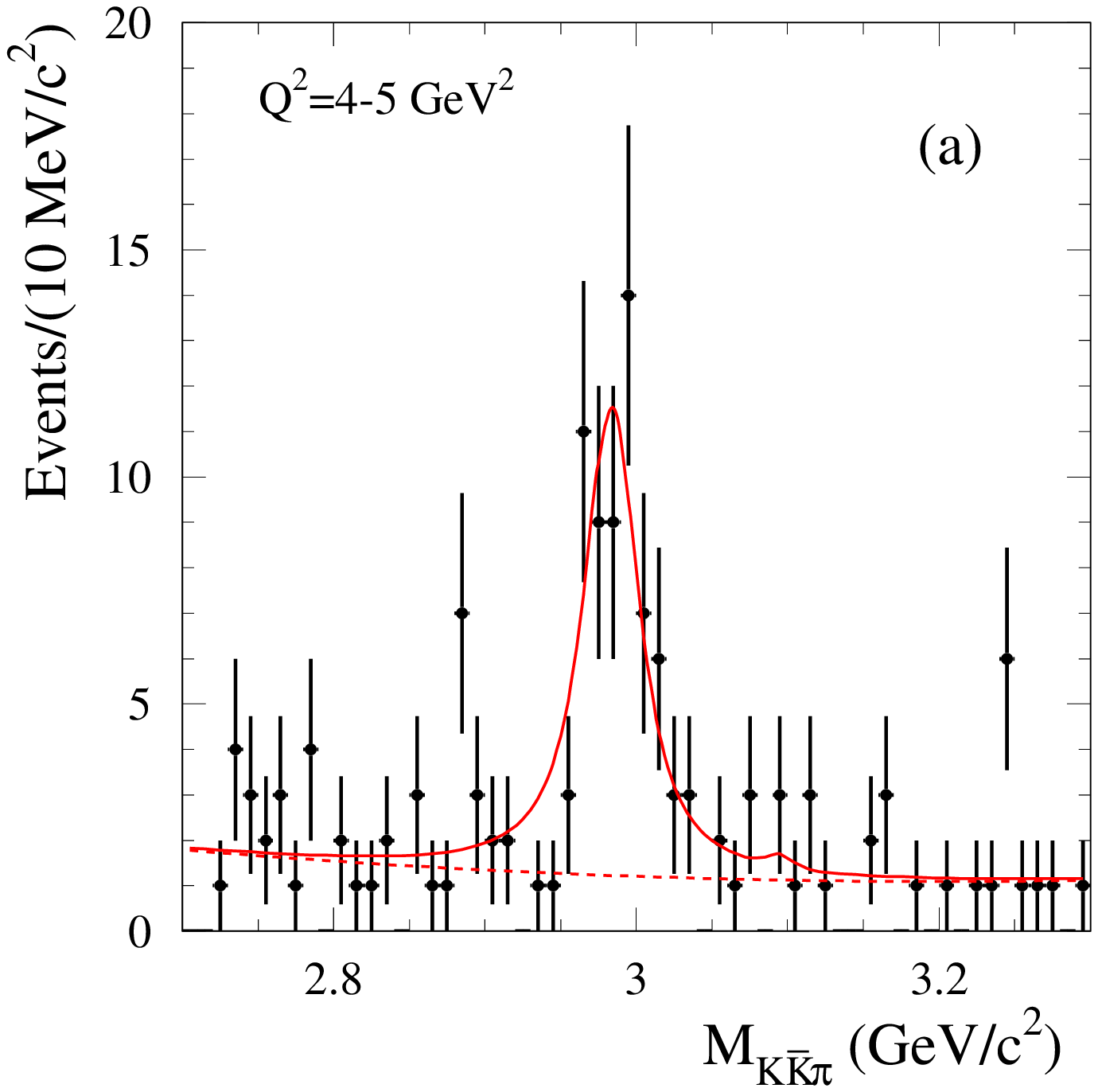}
\includegraphics[width=.32\textwidth]{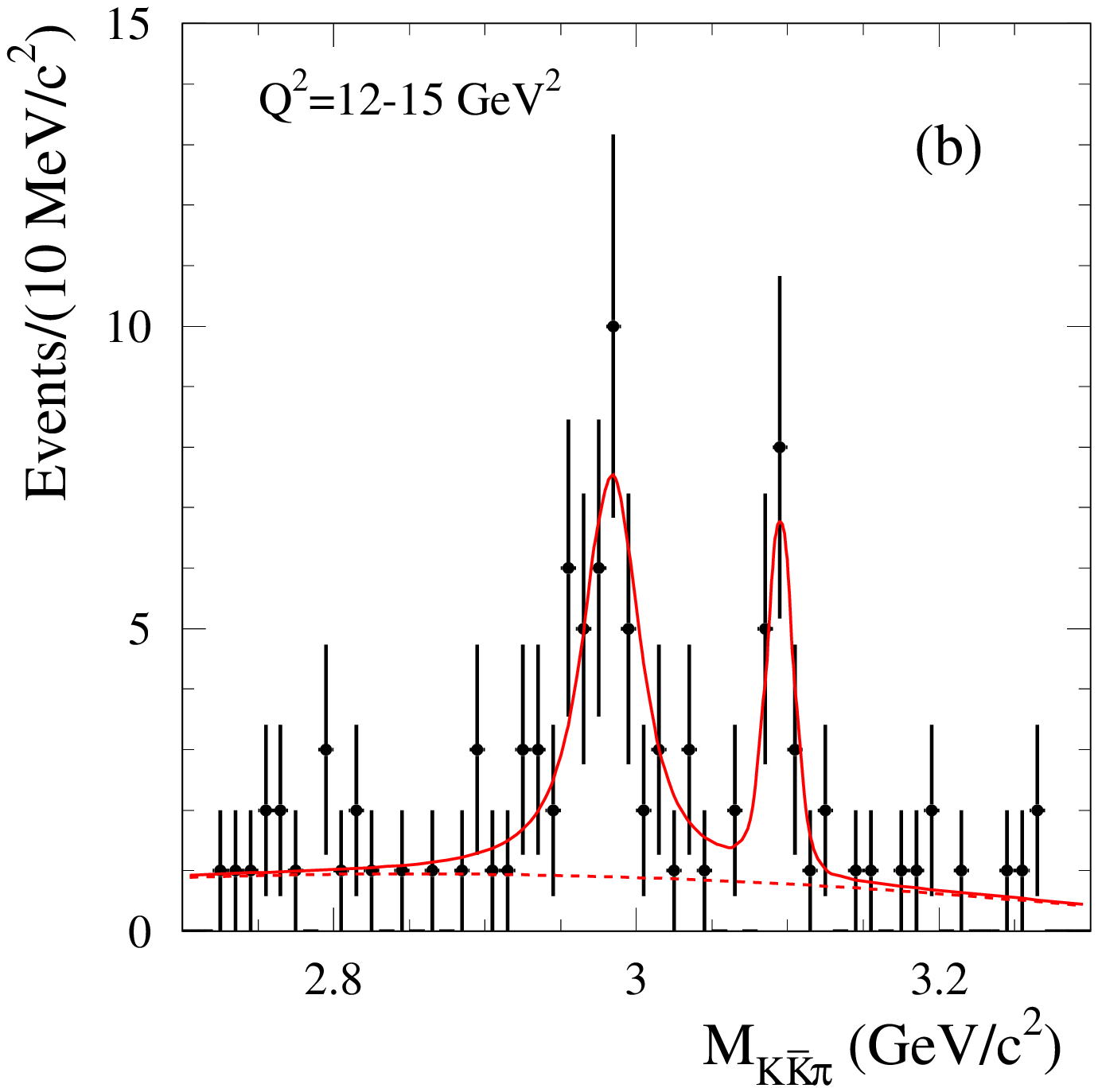}
\includegraphics[width=.32\textwidth]{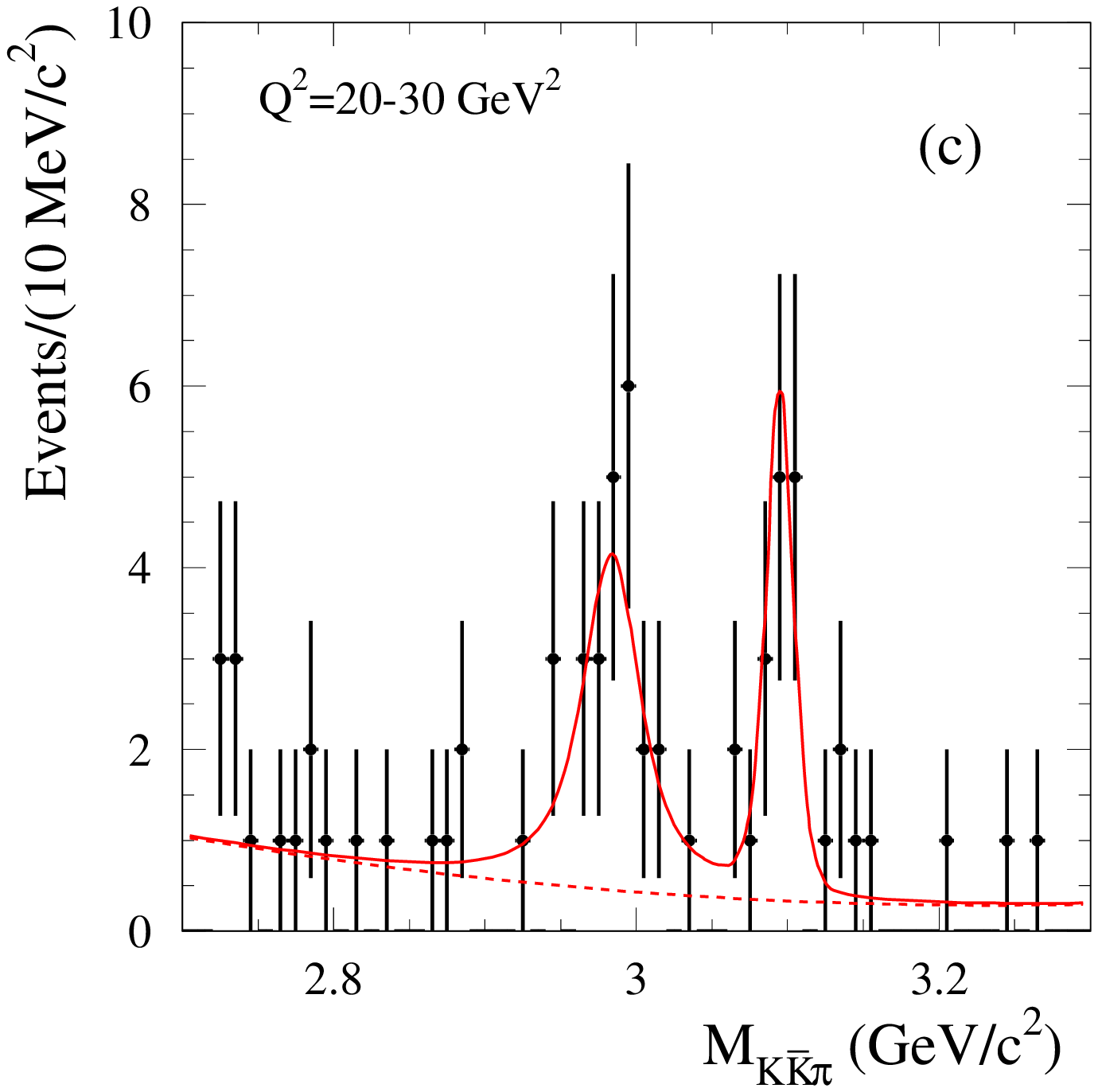}
\caption{The $K_SK^\pm\pi^\mp$ mass spectra for single-tag
data events from three representative $Q^2$ intervals. In each figure the 
solid curve represents the fit result, and the dashed curve indicates 
non-peaking background. \label{fig15}}
\end{figure*}

A fitting procedure similar to that described above is applied in each of 
the eleven $Q^2$
intervals indicated in Table~\ref{tab10}. The parameters of the mass
resolution function are taken from the fit to the mass spectrum for
simulated events in the corresponding $Q^2$ interval. The $\eta_c$ and $J/\psi$
masses are fixed at the values obtained from the fit to the spectrum of
Fig.~\ref{fig14}, while the $\eta_c$  width is taken from the fit to the 
no-tag data.
The free parameters in the fit are the numbers of $\eta_c$ and 
$J/\psi$ events, and two or three additional parameters depending upon the 
description of the background shape (quadratic or exponential). The 
$K_SK^\pm\pi^\mp$ mass spectra 
and fitted curves (quadratic background) for three representative $Q^2$ 
intervals are shown in Fig.~\ref{fig15}.
The number of $\eta_c$ events  obtained from the fit using a
quadratic background is listed for each $Q^2$ interval in Table~\ref{tab10}. 
The difference between the fit results for the two 
background hypotheses is used as an estimate of the systematic uncertainty 
associated with the assumed background form. 
\begin{table*}[t]
\caption{The $Q^2$ interval, number of events with $\eta_c$ obtained from the
fit ($N_{\eta_c}$), number of background events from $J/\psi\to\eta_c\gamma$ 
decay ($N_{\rm bkg}$), efficiency correction ($\delta_{\rm total}$), and number of 
signal events corrected for data/MC difference and resolution effects 
($N_{\rm corr}^{\rm unfolded}$). The quoted errors on $N_{\eta_c}$ and
$N_{\rm corr}^{\rm unfolded}$ are statistical and systematic.
\label{tab10}}
\begin{ruledtabular}
\begin{tabular}{cccccc}
$Q^2$ interval (GeV$^2$) & $N_{\eta_c}$ & $N_{\rm bkg}$ & $\delta_{\rm total}(\%)$ &
$N_{\rm corr}^{\rm unfolded}$ & $\varepsilon (\%)$ \\
\hline 
 no-tag     & $14450\pm320\pm400$ & $730\pm240$  & $-2.6$ & $14090\pm330\pm480$ & 14.5 \\
 2--3  & $41.0\pm8.6\pm1.3$  & $0.7\pm0.4$  & $-0.1$ & $39.9\pm9.0\pm1.4$  &  2.0 \\
 3--4  & $56.2\pm10.5\pm4.0$ & $0.6\pm0.4$  & $ 0.0$ & $55.3\pm10.9\pm4.2$ &  4.9 \\
 4--5  & $65.0\pm10.9\pm1.1$ & $0.1\pm0.4$  & $-0.1$ & $64.8\pm11.5\pm1.2$ &  9.1 \\
 5--6  & $52.6\pm9.6\pm0.6$  & $0.5\pm0.4$  & $-0.4$ & $51.8\pm10.3\pm0.8$ & 12.1 \\
 6--8  & $90.9\pm12.2\pm4.6$ & $1.3\pm0.8$  & $-0.4$ & $90.3\pm12.8\pm4.9$ & 14.0 \\
 8--10 & $60.9\pm10.9\pm2.8$ & $0.9\pm0.6$  & $-0.8$ & $61.3\pm11.7\pm3.1$ & 17.9 \\
10--12 & $34.8\pm7.3\pm1.8$  & $1.0\pm0.6$  & $-1.0$ & $33.5\pm7.9\pm2.1$  & 21.4 \\
12--15 & $42.3\pm8.7\pm2.1$  & $1.9\pm0.8$  & $-1.3$ & $41.2\pm9.4\pm2.4$  & 23.0 \\
15--20 & $45.5\pm7.9\pm1.0$  & $2.4\pm1.0$  & $-1.0$ & $44.3\pm8.5\pm1.5$  & 23.8 \\
20--30 & $23.7\pm6.6\pm0.6$  & $1.6\pm0.7$  & $-1.0$ & $22.5\pm6.9\pm1.0$  & 24.7 \\
30--50 & $10.8\pm4.5\pm0.1$  & $0.9\pm0.5$  & $-1.3$ & $10.3\pm4.8\pm0.5$  & 24.5 \\
\end{tabular}
\end{ruledtabular}
\end{table*}

\section{Peaking background estimation and subtraction}\label{background}
Background containing true $\eta_c$'s might arise from $e^+e^-$ annihilation 
processes and two-photon processes with higher multiplicity
final states. The processes with a $J/\psi$ in the final state considered
in previous sections are also sources of peaking background because
of the relatively large branching fraction for the 
decay $J/\psi\to\eta_c\gamma$~\cite{pdg}.
\begin{figure}
\includegraphics[width=.4\textwidth]{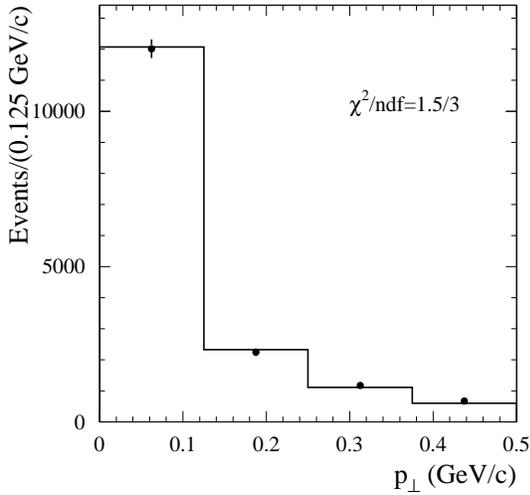}
\caption{The distribution of the $\eta_c$ candidate transverse momentum 
for no-tag data events containing an $\eta_c$ (points with error bars),
and for simulated signal events (histogram).
\label{fig16}}
\end{figure}

For no-tag events the 
most discriminating variable between signal and background
is the $\eta_c$
candidate transverse momentum ($p_\perp^\ast$). In particular it is expected
that background from $e^+e^-$ annihilation increases rapidly with transverse 
momentum. Figure~\ref{fig16} shows the $p_\perp^\ast$ distribution for no-tag 
data events containing an $\eta_c$. In each $p_\perp^\ast$ interval the number
of $\eta_c$ events is determined from the fit to the $K_SK^\pm\pi^\mp$ mass 
spectrum. It is seen that the data distribution is in good agreement with 
signal MC simulation. A conservative upper limit on the level of $e^+e^-$
annihilation background is obtained by fitting a sum of the MC signal 
distribution and a constant background to the data $p_\perp^\ast$ distribution.
The number of background events with $p_\perp^\ast < 0.25$ GeV/$c$ is found to
be $110\pm150$.

\begin{figure}
\includegraphics[width=.4\textwidth]{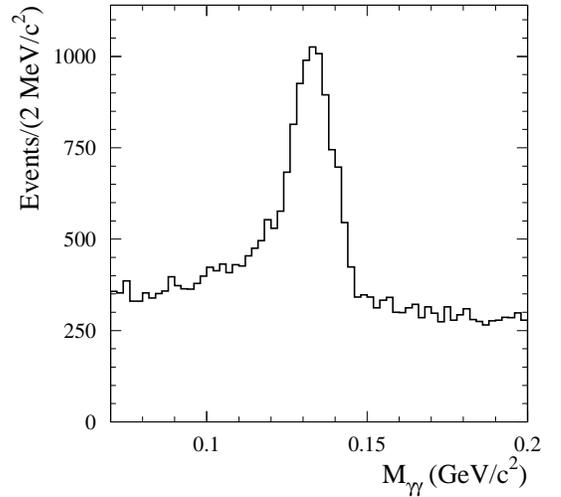}
\caption{The $\pi^0$ candidate mass spectrum for the selected 
$e^+e^-\to e^+e^-K_SK^\pm\pi^\mp\pi^0$ data events in no-tag mode. 
\label{fig17}}
\end{figure}
\begin{figure}
\includegraphics[width=.4\textwidth]{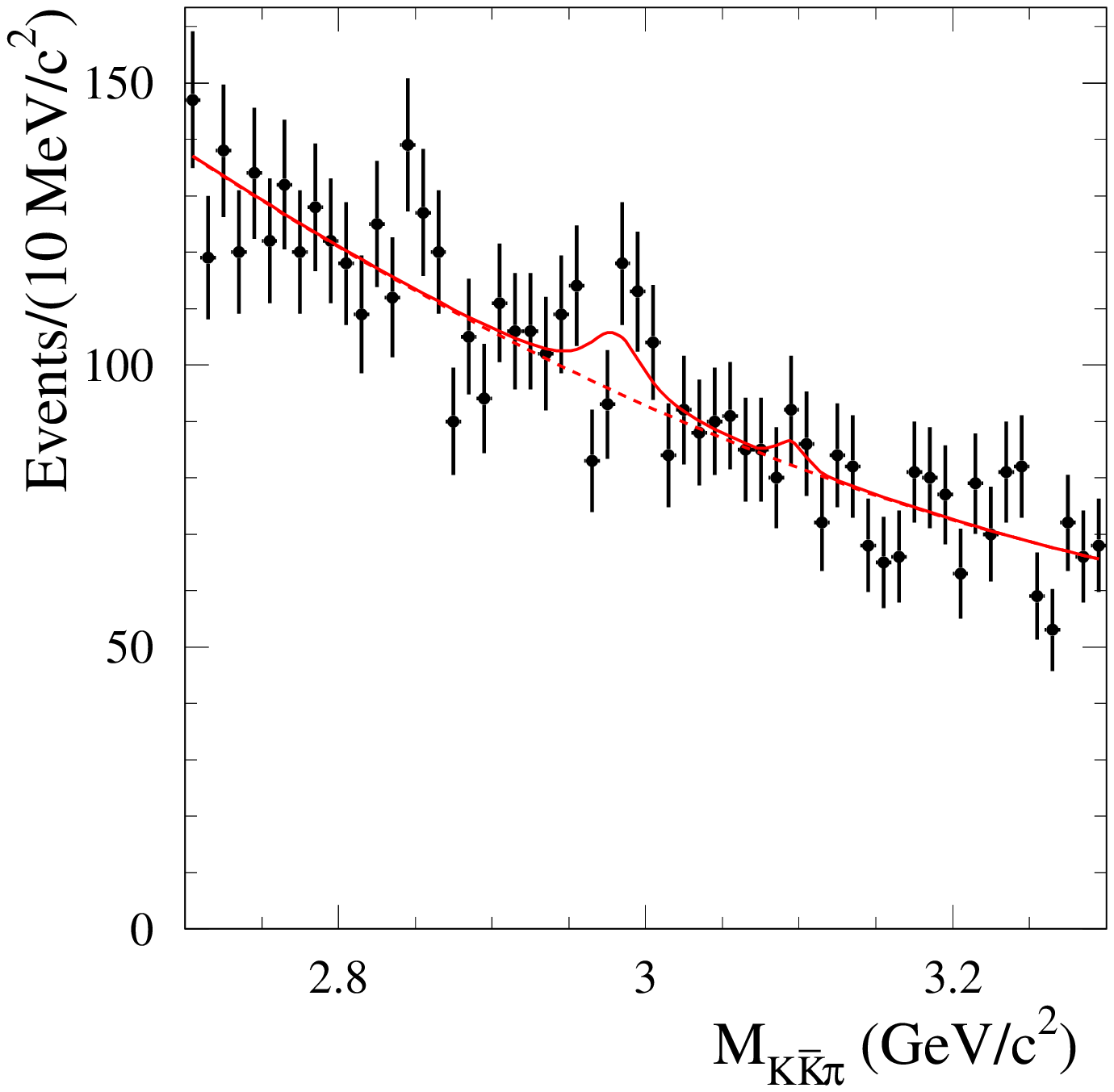}
\caption{The $K_SK^\pm\pi^\mp$ mass spectrum for 
$e^+e^-\to e^+e^-K_SK^\pm\pi^\mp\pi^0$ candidate events 
in no-tag mode with $0.11<M_{\gamma\gamma}<0.15$ GeV/$c^2$. 
The solid curve is the
fit result. The dashed curve represents non-peaking background.
\label{fig18}}
\end{figure}
The two-photon background from the process $e^+e^-\to e^+e^-\eta_c\pi^0$
is studied using a special selection. From the sample of events satisfying
preliminary selection criteria, events with two or more extra photons are 
selected with the energy of each photon required to be greater than 
50 MeV. Two photons with total energy greater than 0.2 GeV, and invariant 
mass in the range 0.07--0.20 GeV/$c^2$ form a $\pi^0$ candidate. The $\pi^0$ 
candidate is combined with an $\eta_c$ candidate, and it is required that the 
transverse momentum of the $\eta_c\pi^0$ system be less than 0.25 GeV/$c$.
The resulting $\pi^0$ candidate mass spectrum is shown in Fig.~\ref{fig17},
and the $K_SK^\pm\pi^\mp$ mass spectrum for events with 
$0.11<M_{\gamma\gamma}<0.15$ GeV/$c^2$ is shown in Fig.~\ref{fig18}. The 
spectrum is fitted by a sum of signal ($\eta_c$ and $J/\psi$) and background 
functions. The fitted number of $\eta_c$ events is found to be $60\pm40$.
Simulation of the $e^+e^-\to e^+e^-\eta_c\pi^0$
process shows that the ratio of the numbers of events selected using
the standard and special criteria is about 2.5, so that
the estimated background from $e^+e^-\to e^+e^-\eta_c\pi^0$ in the event 
sample with the standard selection is $150\pm100$.
A similar approach is used to estimate a possible background from
the $e^+e^-\to e^+e^-\eta_c\eta$ process.  
No $\eta$ signal is observed in the two-photon mass spectrum, nor is there
an $\eta_c$ signal in the $K_SK^\pm\pi^\mp$ mass spectrum.  
The $e^+e^-\to e^+e^-\eta_c\eta$ background is therefore considered negligible.

Background from both sources, $e^+e^-$ annihilation and two-photon
processes, does not exceed 490 events (90\% CL), i.e., 3.5\% of the total 
number of $\eta_c$ events. This number is considered to provide an estimate 
of the systematic uncertainty due to possible $e^+e^-$ annihilation and 
two-photon background.

The total number of $e^+e^-\to J/\psi\gamma,\,J/\psi\to K_SK^\pm\pi^\mp$ events
found in the no-tag event sample is $3170\pm 100$. From simulation the ratio 
of the detection efficiencies
$\varepsilon(e^+e^-\to J/\psi\gamma,\,J/\psi\to\eta_c\gamma,\,\eta_c\to K_SK^\pm\pi^\mp)/
\varepsilon(e^+e^-\to J/\psi\gamma,\,J/\psi\to K_SK^\pm\pi^\mp)$ is found to
be $1.18\pm0.01$. Taking into account the ratio of the branching fractions
\begin{eqnarray}
b=\frac{{\cal B}(J/\psi\to\eta_c\gamma){\cal B}(\eta_c\to K\bar{K}\pi)}
{{\cal B}(J/\psi\to K\bar{K}\pi)}=0.20\pm 0.07
\label{brrat}
\end{eqnarray}
the corresponding background contribution to the $\eta_c$ peak from the ISR 
process is found to be $730\pm240$ events. 

For single-tag events, the background from $e^+e^-$ annihilation can
be estimated using events with the wrong sign of the $e^\pm\eta_c$
momentum  $z$-component.
The $K_SK^\pm\pi^\mp$ mass spectrum for the  wrong-sign events is
shown in Fig.~\ref{fig7} together with the spectrum for 
right-sign events. Assuming that the numbers of background events
for the wrong- and right-sign data samples are approximately
the same, a fit to the mass spectrum for wrong-sign events yields $1.4\pm3.0$
$e^+e^-$ annihilation events peaking at the $\eta_c$ mass.

To estimate the two-photon background in the single-tag event sample,
the result obtained for no-tag events is used. Assuming that the signal and 
background $Q^2$ dependences are approximately the same, it is
estimated that the number of two-photon background events is $5.7\pm3.0$.
The total background from $e^+e^-$ annihilation and two-photon processes
does not exceed 13 events (90\% CL), or 2.5\% of the total number of
$\eta_c$ events in the single-tag event sample. This number is considered
to provide an estimate of the systematic uncertainty due to  possible 
$e^+e^-$ annihilation and two-photon background.

The background from the process
$e^+e^-\to e^+e^-J/\psi$ (Fig.~\ref{fig2}(b)), $J/\psi\to \eta_c\gamma$, 
$\eta_c\to K_SK^\pm\pi^\mp$
is estimated from the measured $Q^2$ distribution ($N_{J/\psi,i}$) for the
$e^+e^-\to e^+e^-J/\psi,\,J/\psi\to K_SK^\pm\pi^\mp$ events as 
$\kappa_i b N_{J/\psi,i}$, where $b$ is the ratio of the branching fractions 
defined in Eq.(\ref{brrat}), and $\kappa_i$ is the ratio of the detection efficiencies
for the $J/\psi\to\eta_c\gamma,\,\eta_c\to K_SK^\pm\pi^\mp$ and 
$J/\psi\to K_SK^\pm\pi^\mp$ decay modes. The coefficient $\kappa_i$ varies 
from 0.7 to 0.5 in the $Q^2$ range of interest. The estimated 
background contributions resulting from $J/\psi\to \eta_c\gamma$ decay 
are listed in Table~\ref{tab10}. The fraction of background
in the $e^\pm\eta_c$ data sample changes from about 1.0\% for 
$Q^2<10$ GeV$^2$ to about 5\% at $Q^2\approx 30$ GeV$^2$.

\section{Detection efficiency}
\begin{figure}
\includegraphics[width=.4\textwidth]{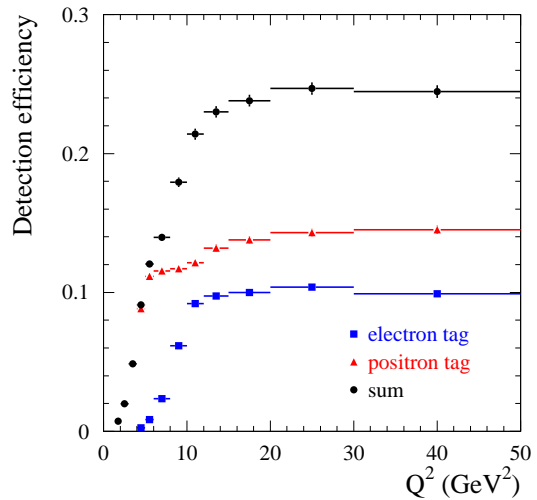}
\caption{The detection efficiency as a function of momentum transfer
squared for events with a tagged electron (squares), a tagged positron
(triangles), and their sum (circles).
\label{fig19}}
\end{figure}
\begin{figure}
\includegraphics[width=.33\textwidth]{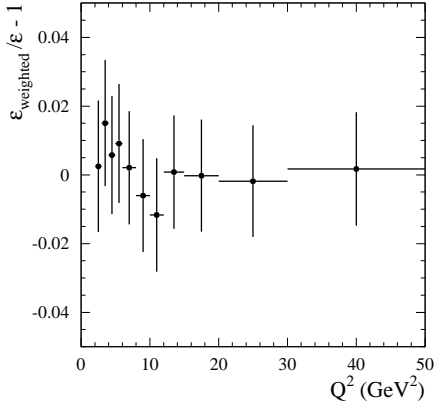}
\caption{The $Q^2$ dependence of the relative difference between detection
efficiencies determined from MC simulation with and without Dalitz-plot 
reweighting.
\label{fig20}}
\end{figure}
The detection efficiency is determined from MC simulation as the ratio
of the true $Q^2$ distributions computed after and before applying the
selection criteria. The $Q^2$ dependence of the detection efficiency
is shown in Fig.~\ref{fig19}. The detector acceptance limits the detection
efficiency at small $Q^2$. The cross section is measured in the region 
$Q^2 > 2$ GeV$^2$ where the detection efficiency is greater than 2\%. 
The asymmetry of the $e^+e^-$ collisions at PEP-II leads to different 
efficiencies for events with electron and positron tags. The $Q^2$ range 
from 2 to 6 GeV$^2$ is measured only with the positron tag. For no-tag events
the detection efficiency is $0.1446\pm0.0023$. The efficiency is calculated 
using simulated events reweighted according to the Dalitz plot distribution 
observed in data. 
For no-tag events the relative difference 
between detection efficiencies calculated with and without weighting is found 
to be $-(1.1\pm1.6)\%$. The quoted error is determined by the statistical 
errors of the measured Dalitz plot distribution. The corresponding relative
difference for single-tag events is shown in Fig.~\ref{fig20} as a function 
of $Q^2$. The detection efficiency has only a very weak dependence on the 
dynamics of $\eta_c$ decay.

\begin{figure}
\includegraphics[width=.33\textwidth]{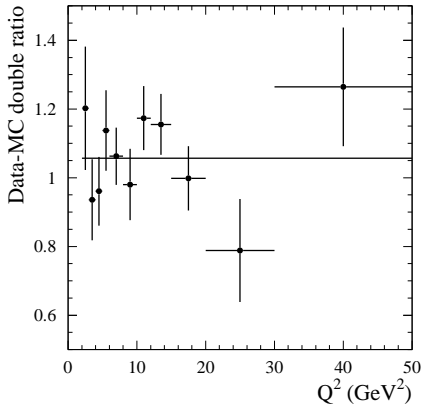}
\caption{The $Q^2$ dependence of the data-MC double ratio for events
with and without the polar angle restriction $0.387<\theta<2.400$ rad for
all four charged-particle tracks from the $\eta_c$ decay.
\label{fig21}}
\end{figure}
Possible sources of systematic uncertainty due to differences
between data and simulated detector response are now considered. For no-tag 
events the MC simulation predicts a significant loss of signal events,
$(5.3\pm0.1)$\%, due to background filters used in event 
reconstruction. The filter inefficiency can be measured in data using
a special sample of prescaled events that does not pass the background 
filters. The filter inefficiency obtained in data is $(7.5\pm1.2)$\%. 
The difference $\delta=-(2.2\pm1.2)$\% is used to correct the number of signal
events. For single-tag events the presence of the additional electron
leads to a significantly smaller filter inefficiency. The simulation predicts 
a filter inefficiency of $(0.57\pm0.02)$\%, which is about ten times smaller 
than for no-tag events. We conclude that this source of systematic uncertainty
is negligible for single-tag events. 

\begin{figure}
\includegraphics[width=.33\textwidth]{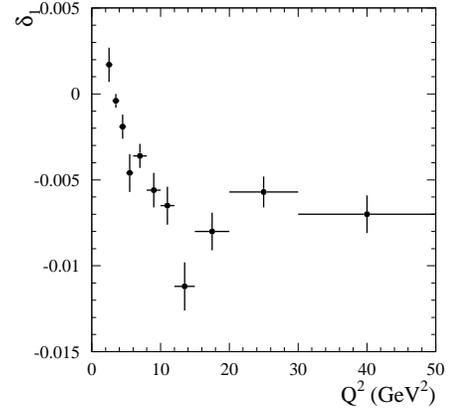}
\caption{ The correction to the MC-estimated kaon identification efficiency
as a function of $Q^2$.
\label{fig22}}
\end{figure}
\begin{figure}
\includegraphics[width=.33\textwidth]{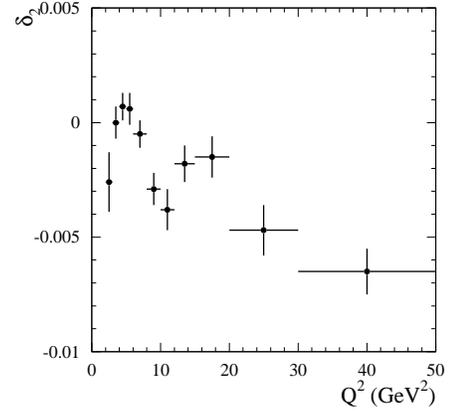}
\caption{The correction to the MC-estimated electron identification
efficiency as a function of $Q^2$.
\label{fig23}}
\end{figure}
To study the possible systematics for no-tag events due to selection criteria,
the mass window for the $K_S$ candidate is increased 
from 487.5--507.5 MeV/$c^2$ to 475.0--520.0 MeV/$c^2$, the limit on the
$\eta_c$ candidate transverse momentum is changed from 0.25 to 0.50 GeV/$c$,
and the polar angle restriction is set to $0.387<\theta<2.400$ rad for all four 
charged-particle tracks in an event. The last modification rejects about 30\% 
of signal events. The double ratio
\begin{equation}
R_2=\frac{(N_{\rm new}/N)_{\rm data}}{(N_{\rm new}/N)_{\rm MC}}
\label{drat}
\end{equation}
is calculated,
where  $N_{\rm new}$ and $N$ are the numbers of signal events with the new
and standard selection criteria, and is found to be close to unity
for the definition of the $K_S$ mass window ($0.993\pm0.005$), and for
the condition on $p_\perp^\ast$ ($1.002\pm0.009$). A significant deviation
from unity $(5.9\pm 1.8)\%$ is observed for the polar angle restriction. 
This deviation is taken as an estimate of the systematic uncertainties due to
imperfect simulation of the $\eta_c$ selection criteria close to the limits
of the fiducial tracking region.

The systematic uncertainty due to a possible difference between the data and
simulation in the charged-particle track reconstruction for pions and kaons 
is estimated to be about 
0.35\% per track, so that the total uncertainty is estimated to be 1.4\%. 
The data-MC simulation 
difference in the kaon identification is estimated using the identification 
efficiencies measured for kaons from the 
$D^{\ast +}\to D^0\pi^+,\;D^0\to\pi^+ K^-$ decay sequence. The
efficiency correction is found to be $-0.4\%$. The systematic uncertainty
associated with this correction is taken to be equal to the value of 
the correction, 0.4\%. The total efficiency correction for no-tag events
is $-2.6\%$, and the total systematic error associated with the efficiency is 
about 6.2\%.

The polar angle restriction described above is also tested for single-tag
events, and Fig.~\ref{fig21} shows the $Q^2$ dependence of the data-MC double
ratio.  No significant $Q^2$ dependence is observed, and the 
average value is $1.057\pm0.032$, which is very close to the value
for no-tag events. For single-tag events the systematic uncertainty due to 
the $\eta_c$ selection criteria is estimated to be 5.7\%. 
The efficiency correction
($\delta_1$) for kaon identification for single-tag events is shown in 
Fig.~\ref{fig22} as a function of $Q^2$, and this results in an associated
systematic uncertainty of 0.5\%.

\begin{table}[t]
\caption{The main sources of systematic uncertainty associated with the
detection efficiency and the total efficiency systematic error for no-tag
and single-tag events.
\label{tab20}}
\begin{ruledtabular}
\begin{tabular}{lcc}
Source                   & No-tag, \% & Single-tag, \% \\
\hline
trigger, filters         & 1.2        & --             \\
$\eta_c$ selection       & 5.9        & 5.7            \\
track reconstruction     & 1.4        & 1.5            \\
$K^\pm$ identification   & 0.4        & 0.5            \\ 
$e^\pm$ identification   & --         & 0.5            \\
\hline
total                    & 6.2        & 5.9            \\
\end{tabular}
\end{ruledtabular}
\end{table}
The data-MC simulation difference in electron identification is estimated
using the identification efficiencies measured for electrons from 
radiative Bhabha events. The efficiency correction ($\delta_2$) 
is shown as a function of $Q^2$  in Fig.~\ref{fig23}.
The associated systematic uncertainty
does not exceed 0.5\%. The systematic uncertanty due to data-MC simulation
difference in the electron track reconstruction is about 0.1\%.
To estimate the effect of the requirement $-0.02<r<0.03$ (see
Fig.~\ref{fig6}),
events with $0.03<r<0.06$ are studied. The data-MC simulation double ratio 
defined by Eq.(\ref{drat}) is found to be consistent with unity, ($0.99\pm0.02$), 
so that 
the simulation reproduces the shape of the $r$ distribution very well. 

The main sources of systematic uncertainty associated with detection efficiency
are summarized in Table~\ref{tab20} for the no-tag and single-tag samples.
The values of the detection efficiency and total efficiency correction
$\delta_{\rm total,i}\approx\delta_{1,i}+\delta_{2,i}$ for different $Q^2$ 
intervals are listed in Table~\ref{tab10}. The data distribution is corrected 
as follows:
\begin{equation}
N_{{\rm corr},i}=N_i/(1+\delta_{{\rm total},i}),
\label{eqcor}
\end{equation}
where $N_i=N_{\eta_c,i}-N_{{\rm bkg},i}$ is the number of signal events in 
the $i$th $Q^2$ interval. 

\section{Cross section and form factor}\label{crosssec}
The Born differential cross section for 
$e^+e^-\to e^+e^-\eta_c, \eta_c \to K\bar{K}\pi$ is 
\begin{multline}
\frac{{\rm d}\sigma}{{\rm d}Q^2}=
\frac{({\rm d}N/{\rm d}Q^2)_{\rm corr}^{\rm unfolded}}{\varepsilon RL}\times\\
\frac{{\cal B}(\eta_c \to K\bar{K}\pi)}{{\cal B}(\eta_c \to K_S K^\pm\pi^\mp)
{\cal B}(K_S\to\pi^+\pi^-)},
\label{eqcs}
\end{multline}
where $({\rm d}N/{\rm d}Q^2)_{\rm corr}^{\rm unfolded}$ is the mass spectrum 
corrected for 
data-MC simulation difference and unfolded for detector resolution effects,
explained below,
$L$ is the total 
integrated luminosity, 
$\varepsilon$ is the $Q^2$-dependent 
detection efficiency, $R$ is a radiative correction factor 
accounting for distortion of the $Q^2$ spectrum due to the emission of 
photons from the initial-state particles, and for vacuum polarization effects. 
The ratio ${\cal B}(\eta_c \to K\bar{K}\pi)/{\cal B}(\eta_c \to K_S K^\pm\pi^\mp)$ is
expected to be 3 from isospin relations.

\begin{figure}
\includegraphics[width=.33\textwidth]{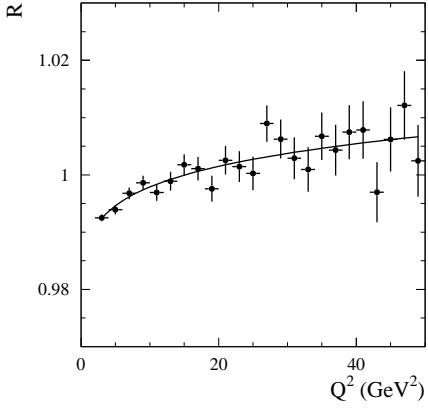}
\caption{The $Q^2$ dependence of the radiative correction factor.
\label{fig24}}
\end{figure}
The radiative correction factor is determined using simulation
at the generator level (with no detector simulation). The $Q^2$ spectrum
is generated using only the pure Born amplitude for the 
$e^+e^-\to e^+e^-\eta_c$ process, and then using a model with radiative 
corrections included. The $Q^2$ dependence of the radiative correction factor, 
evaluated as the ratio of the second spectrum to the first, is shown in 
Fig.~\ref{fig24}, and is fitted using the function $a/(1+bQ^\gamma)$.
The accuracy of the radiative correction calculation is estimated to be 
1\%~\cite{RC}.
It should be noted that the value of $R$ depends on the requirement on
the extra photon energy. The $Q^2$ dependence obtained corresponds to
the condition $r=2E^\ast_\gamma/\sqrt{s}<0.1$ imposed in the simulation.

The corrected and unfolded $Q^2$ distribution 
$({\rm d}N/{\rm d}Q^2)_{\rm corr}^{\rm unfolded}$ 
is obtained from the measured distribution 
by dividing by the efficiency correction factor (see Eq.(\ref{eqcor})) and 
unfolding the effect of $Q^2$ resolution. Using MC simulation,
a migration matrix $H$ is obtained, which represents the probability that
an event with true $Q^2$ in interval $j$ be reconstructed in interval $i$:
\begin{equation}
\left( \frac{{\rm d}N}{{\rm d}Q^2} \right)^{\rm rec}_i=
\sum_{j}H_{ij}\left( \frac{{\rm d}N}{{\rm d}Q^2} \right)^{\rm true}_j.
\end{equation}
In the case of extra photon emission, $Q^2_{\rm true}$ is calculated
as $-(p-p^\prime-k)^2$, where $k$ is the photon four-momentum;
$\varepsilon$ and $R$ in Eq.(\ref{eqcs}) are functions of $Q^2_{\rm true}$.
As the chosen $Q^2$ interval width significantly exceeds the resolution
for all $Q^2$, the migration matrix is nearly diagonal, with values
of diagonal elements $\sim 0.95$, and of the next-to-diagonal $\sim 0.02$.
The true $Q^2$ distribution is obtained by applying the inverse of the 
migration matrix to the measured distribution. The procedure changes the shape
of the $Q^2$ distribution insignificantly, but increases the errors
(by $\approx$10\%) and their correlations. The corrected $Q^2$ spectrum 
($N_{\rm corr}^{\rm unfolded}$) is listed in Table~\ref{tab10}.
  
\begin{figure}[t]
\includegraphics[width=.4\textwidth]{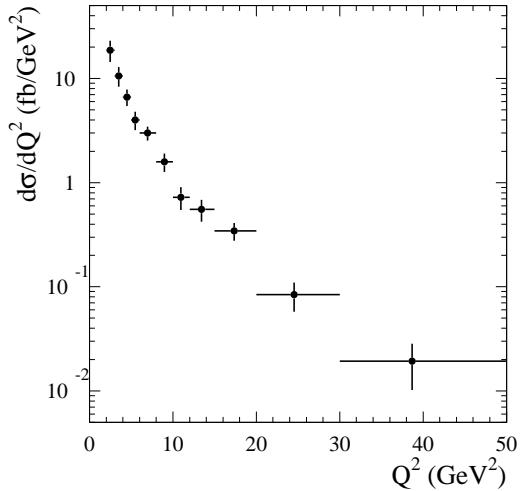}
\caption{The $e^+e^-\to e^+e^-\eta_c$ differential cross section
multiplied by the $\eta_c\to K\bar{K}\pi$ branching fraction;
the statistical and $Q^2$-dependent systematic errors of Table~\ref{tab30}
have been combined in quadrature.
\label{fig25}}
\end{figure}

The values of the differential cross sections obtained are listed
in Table~\ref{tab30}, where the first error is statistical and
the second systematic.
The latter includes only $Q^2$-dependent errors, namely, the 
systematic uncertainty in the number of signal events and the statistical 
errors on the efficiency correction and MC simulation. The $Q^2$-independent
systematic error is 6.6\%; this results from the systematic uncertainties on
the detection efficiency (5.9\%), background subtraction (2.5\%), the radiative
correction factor (1\%), and the error on the integrated luminosity (1\%) 
combined in quadrature.
The MC simulation for single-tag events is performed, and the detection 
efficiency is determined, with the restriction that the momentum transfer
to the untagged electron be greater than $-1$ GeV$^2$, so that the cross section
is measured for the restricted range $|q^2| < 1$ GeV$^2$.
The measured differential cross section is shown in Fig.~\ref{fig25}.
\begin{table}[t]
\caption{The $Q^2$ interval and the weighted average $Q^2$ value 
($\overline{Q^2}$), the $e^+e^-\to e^+e^-\eta_c$ cross section
multiplied by ${\cal B}(\eta_c\to K\bar{K}\pi)$ 
$[{\rm d}\sigma/{\rm d}Q^2(\overline{Q^2})]$, and the normalized
$\gamma\gamma^\ast\to \eta_c$ transition form factor
($|F(\overline{Q^2})/F(0)|$). The statistical and
systematic errors are quoted separately for the cross section, but are combined
in quadrature for the form factor.
Only $Q^2$-dependent systematic errors are quoted; the
$Q^2$-independent error is 6.6\% for the cross section and
4.3\% for the form factor.
\label{tab30}}
\begin{ruledtabular}
\begin{tabular}{cccccc}
$Q^2$ interval & $\overline{Q^2}$ &
${\rm d}\sigma/{\rm d}Q^2(\overline{Q^2})$ &
$|F(\overline{Q^2})/F(0)|$ \\
(GeV$^2$)      &     (GeV$^2$)    &                (fb/GeV$^2$)     &  \\
\hline 
 2--3  & 2.49  & $18.7\pm 4.2\pm 0.8$    & $0.740\pm0.085$ \\
 3--4  & 3.49  &$10.6\pm 2.1\pm 0.8$     & $0.680\pm0.073$ \\
 4--5  & 4.49  &$6.62\pm 1.18\pm 0.19$   & $0.629\pm0.057$ \\
 5--6  & 5.49  &$4.00\pm 0.80\pm 0.10$   & $0.555\pm0.056$ \\
 6--8  & 6.96  &$3.00\pm 0.43\pm 0.17$   & $0.563\pm0.043$ \\
 8--10 & 8.97  &$1.58\pm 0.30\pm 0.08$   & $0.490\pm0.049$ \\
10--12 & 10.97 &$0.72\pm 0.17\pm 0.05$   & $0.385\pm0.048$ \\
12--15 & 13.44 &$0.55\pm 0.13\pm 0.03$   & $0.395\pm0.047$ \\
15--20 & 17.35 &$0.34\pm 0.07\pm 0.01$   & $0.385\pm0.038$ \\
20--30 & 24.53 &$0.084\pm0.026\pm 0.004$ & $0.261\pm0.041$ \\
30--50 & 38.68 &$0.019\pm 0.009\pm 0.001$& $0.204\pm0.049$ \\
\end{tabular}
\end{ruledtabular}
\end{table}

Because of the strong nonlinear dependence of the cross section on $Q^2$, 
the value of $Q^2$ corresponding to the measured cross section differs
slightly from 
the center of the $Q^2$ interval. The measured cross section is described by
a smooth function, which is then used to reweight the simulated $Q^2$ distribution 
and calculate the weighted average value ($\overline{Q^2}$) for each $Q^2$ 
interval. The values of $\overline{Q^2}$ obtained are listed in 
Table~\ref{tab30}.

The no-tag event sample is used to obtain the total cross section for the 
reaction $e^+e^-\to e^+e^-\eta_c,\; \eta_c\to K\bar{K}\pi$: 
\begin{multline}
\sigma_{\rm total}=\frac{N_{\rm corr}}{\varepsilon L}\times\\
\frac{{\cal B}(\eta_c \to K\bar{K}\pi)}{{\cal B}(\eta_c \to K_S K^\pm\pi^\mp)
{\cal B}(K_S\to\pi^+\pi^-)}.
\label{xstot}
\end{multline}
In the no-tag mode the radiative correction is expected to be less than 
1\%~\cite{ntRC}, and so the associated systematic uncertainty is
assigned conservatively as 1\%.
Taking $N_{\rm corr}$ and $\varepsilon$ from Table~\ref{tab10}, the value obtained
from Eq.(\ref{xstot}) is
\begin{equation}
\sigma_{\rm total}=0.900\pm0.021\pm0.074\mbox{ pb}.
\label{sigt}
\end{equation}
The quoted errors are statistical and systematic, respectively.
The latter is 8.1\% and includes the systematic uncertainty in
the number of signal events (3.3\%), the statistical and systematic errors on 
the detection efficiency (1.6\% and 6.2\%, respectively), 
on the background subtraction (3.5\%), on the radiative 
correction (1\%), and the error on the integrated luminosity (1\%).
Using MC simulation, the calculated total cross section corresponding to 
$\Gamma(\eta_c \to \gamma\gamma){\cal B}(\eta_c \to K\bar{K}\pi) = 1$ keV is found to
be 2.402 pb, and hence from Eq.(\ref{sigt}) the value
\begin{multline}
\Gamma(\eta_c \to \gamma\gamma){\cal B}(\eta_c \to K\bar{K}\pi) =\\
0.374\pm0.009\pm0.031\mbox{ keV}
\end{multline}
is obtained.
This result agrees with the Particle Data Group
value $0.44\pm0.05$ keV~\cite{pdg}, 
and also with the recent CLEO measurement $0.407\pm0.022\pm0.028$ 
keV~\cite{CLEO}.

To extract the transition form factor, the measured and 
calculated $Q^2$ spectra are compared. The simulation for single-tag
events uses a constant form factor value, so that the measured normalized
form factor is determined from
\begin{multline}
|F^2(Q^2)/F^2(0)|=\frac{({\rm d}N/{\rm d}Q^2)^{\rm data}_{\mbox{\scriptsize{single-tag}}}}
{N^{\rm data}_{\mbox{\scriptsize{no-tag}}}}\times\\
\frac{\varepsilon_{\mbox{\scriptsize{no-tag}}}\sigma_{\rm total}^{\rm MC}}
{\varepsilon_{\mbox{\scriptsize{single-tag}}}(Q^2)({\rm d}\sigma/{\rm d}Q^2)^{\rm MC}_{\mbox{\scriptsize{single-tag}}}}.
\end{multline}

\begin{figure}[t]
\includegraphics[width=.4\textwidth]{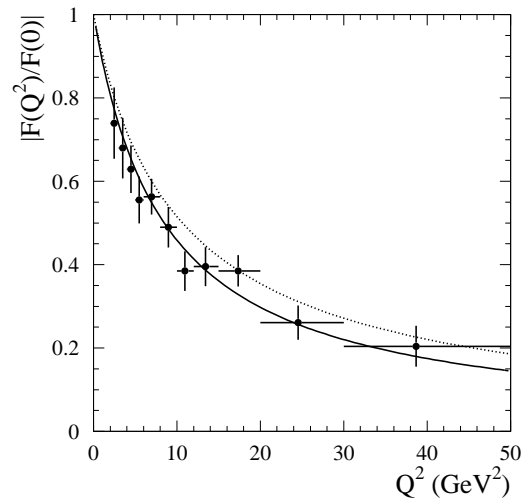}
\caption{The $\gamma\gamma^\ast\to \eta_c$ transition form factor
normalized to $F(0)$ (points with error bars). The solid curve 
shows the fit to Eq.(\ref{mono}).
The dotted curve shows the leading order pQCD prediction from Ref.~\cite{th1}.
\label{fig26}}
\end{figure}
The normalized form factor is proportional to the ratio of the number of 
single-tag events to the number of no-tag events. It is expected that part
of the systematic uncertainty, in particular that associated with the detection
efficiency, cancels in this ratio. However, the single-tag data sample is not
large enough for a detailed study of the $Q^2$-dependence of the 
observed data-MC simulation difference in detector response. Therefore,  
this uncertainty is estimated conservatively to be equal to the corresponding
systematic uncertainty for the single-tag events (6\%). The values of the form
factor obtained are listed in Table~\ref{tab30}, and shown in
Fig.~\ref{fig26}, with the statistical and $Q^2$-dependent systematic errors
combined. 
The $Q^2$-independent systematic error on the form factor is 
4.3\%; this value combines in quadrature the systematic uncertainty on 
detection efficiency, the uncertainty on the number of no-tag events, 
the statistical error on the detection efficiency for no-tag events,
the uncertainties associated with the background subtraction, 
and the uncertainty on the radiative correction.

The form factor data of Fig.~\ref{fig26} are well described 
by the monopole form
\begin{equation}
|F(Q^2)/F(0)|=\frac{1}{1+Q^2/\Lambda},
\label{mono}
\end{equation}
as shown by the solid curve. The corresponding fitted value of the pole
parameter $\Lambda$ is
\begin{equation}
\Lambda=8.5\pm0.6\pm0.7\mbox{ GeV}^2,
\end{equation}
where the second quoted error is due to the 4.3\% $Q^2$-independent systematic
error on the measurements. This value of the pole parameter is in reasonable 
agreement with that expected from vector dominance, namely
$\Lambda=m^2_{J/\psi}=9.6$ GeV$^2$, and in good agreement with 
the lattice QCD calculation, $\Lambda= 8.4\pm0.4$ GeV$^2$~\cite{lqcd}. The dotted curve in
Fig.~\ref{fig26} shows the result of the leading-order pQCD calculation of
Ref.~\cite{th1}. The data lie systematically below this prediction, but within
the theoretical uncertainty quoted in Ref.~\cite{th1}.

\section{Summary}
The reaction $e^+e^- \to e^+e^-\eta_c$, with $\eta_c\to K_S K^\pm\pi^\mp$, 
has been studied
in the no-tag and single-tag modes. 
We measure the following values for the $\eta_c$ mass and width:
\begin{eqnarray}
m_{\eta_c} &=& 2982.2\pm0.4\pm1.6\mbox{ MeV/$c^2$},\nonumber\\
\Gamma &=& 31.7\pm1.2\pm0.8 \mbox{ MeV}.
\end{eqnarray}
These results agree with earlier \babar\ measurements~\cite{bb_etac}
and supersede them.

We have also measured the total cross section
$\sigma(e^+e^- \to e^+e^-\eta_c){\cal B}(\eta_c\to K\bar{K}\pi)$
and differential cross section 
$({\rm d}\sigma/{\rm d}Q^2){\cal B}(\eta_c\to K\bar{K}\pi)$.
From these data we determine the value
\begin{multline}
\Gamma(\eta_c \to \gamma\gamma){\cal B}(\eta_c \to K\bar{K}\pi) =\\
0.374\pm0.009\pm0.031\mbox{ keV}
\end{multline}
and measure the normalized $\gamma\gamma^\ast\to \eta_c$ transition 
form factor $|F(Q^2)/F(0)|$ for the momentum transfer range from 2 to
50 GeV$^2$. The latter is well described by the
simple monopole form of Eq.(\ref{mono}) with
$\Lambda=8.5\pm0.6\pm0.7\mbox{ GeV}^2$ in agreement with both the vector
dominance expectation and the QCD prediction.
\begin{acknowledgments}
We thank V.~L.~Chernyak for useful discussions. We are grateful for the                                                 
extraordinary contributions of our \pep2\ colleagues in                 
achieving the excellent luminosity and machine conditions               
that have made this work possible.                                      
The success of this project also relies critically on the               
expertise and dedication of the computing organizations that            
support \babar.                                                         
The collaborating institutions wish to thank                            
SLAC for its support and the kind hospitality extended to them.         
This work is supported by the                                           
US Department of Energy                                                 
and National Science Foundation, the                                    
Natural Sciences and Engineering Research Council (Canada),             
the Commissariat \`a l'Energie Atomique and                             
Institut National de Physique Nucl\'eaire et de Physique des Particules 
(France), the                                                           
Bundesministerium f\"ur Bildung und Forschung and                       
Deutsche Forschungsgemeinschaft                                         
(Germany), the                                                          
Istituto Nazionale di Fisica Nucleare (Italy),                          
the Foundation for Fundamental Research on Matter (The Netherlands),    
the Research Council of Norway, the                                     
Ministry of Science and Technology of the Russian Federation,           
Ministerio de Educaci\'on y Ciencia (Spain), and the                    
Science and Technology Facilities Council (United Kingdom).             
Individuals have received support from                                  
the Marie-Curie IEF program (European Union) and                        
the A. P. Sloan Foundation.                   
\end{acknowledgments}




\begin{thebibliography}{99}
\bibitem{BKT}
S.~J.~Brodsky, T.~Kinoshita and H.~Terazawa, 
Phys. Rev. D {\bf 4}, 1532 (1971).
\bibitem{LB}
G.~P.~Lepage and S.~J.~Brodsky,
Phys.\ Rev.\  D {\bf 22}, 2157 (1980).
\bibitem{th1}
T.~Feldmann and P.~Kroll,
Phys.\ Lett.\  B {\bf 413}, 410 (1997).
\bibitem{th2}
F.~G.~Cao and T.~Huang,
Phys.\ Rev.\  D {\bf 59}, 093004 (1999).
\bibitem{lqcd}
J.~J.~Dudek and R.~G.~Edwards,
Phys.\ Rev.\ Lett.\  {\bf 97}, 172001 (2006).
\bibitem{L3}
M.~Acciarri {\it et al.}  [L3 Collaboration],
Phys.\ Lett.\  B {\bf 461}, 155 (1999).
\bibitem{pdg}C.~Amsler {\it et al.} (Particle Data Group), 
Phys. Lett. B {\bf 667}, 1 (2008) and 2009 partial update for the 2010
edition (URL:http://pdg.lbl.gov).
\bibitem{babar-nim} 
B.~Aubert {\em et al.} [\babar\ Collaboration],
Nucl. Instr. and Meth. A {\bf 479}, 1 (2002).
\bibitem{RC}
S.~Ong and P.~Kessler,
Phys.\ Rev.\  D {\bf 38}, 2280 (1988).
\bibitem{GEANT4} S.~Agostinelli {\em et al.} [GEANT4 Collaboration],
Nucl. Instr. and Meth. A {\bf 506}, 250 (2003).
\bibitem{bb_etac}B.~Aubert {\it et al.}  [\babar\ Collaboration],
Phys.\ Rev.\ Lett.\  {\bf 92}, 142002 (2004).
\bibitem{ntRC}
M.~Defrise, S.~Ong, J.~Silva and C.~Carimalo,
Phys.\ Rev.\  D {\bf 23}, 663 (1981);
W.~L.~van Neerven and J.~A.~M.~Vermaseren,
Nucl.\ Phys.\  B {\bf 238}, 73 (1984).
\bibitem{CLEO}
D.~M.~Asner {\it et al.}  [CLEO Collaboration],
Phys.\ Rev.\ Lett.\  {\bf 92}, 142001 (2004).
\end{thebibliography}
\end{document}